\documentclass[aps,prb,10pt,twocolumn,amsmath,amssymb,
    longbibliography,superscriptaddress,nofootinbib,
    showpacs,preprintnumbers]{revtex4-1}

\usepackage[colorlinks=true,citecolor=blue,linkcolor=blue]{hyperref}
\usepackage{oubraces}
\usepackage{bm,bbm,mathbbol}
\usepackage{physics}
\usepackage{mathtools}
\usepackage{amsmath, amsthm, amssymb, amsfonts}
\usepackage{bbold}
\usepackage{esvect}

\usepackage{epsfig}
\usepackage{array}
\usepackage{multirow}
\usepackage{graphicx}

\newcommand{\fnorm}[1]{\left|\left|{#1}\right|\right|_{\rm F}}

\begin{document}

\title{Reliable methods for seamless stitching of tight-binding models based on maximally localized Wannier functions}

\author{Jae-Mo Lihm}
\author{Cheol-Hwan Park}
\email{cheolhwan@snu.ac.kr}
\affiliation{Department of Physics, Seoul National University, Seoul 08826, Korea}
\date{\today}

\begin{abstract}
Maximally localized Wannier functions are localized orthogonal functions that can accurately represent given Bloch eigenstates of a periodic system at a low computational cost, thanks to the small size of each orbital. Tight-binding models based on the maximally localized Wannier functions obtained from different systems are often combined to construct tight-binding models for large systems such as a semi-infinite surface. However, the corresponding maximally localized Wannier functions in the overlapping region of different systems are not identical, and this discrepancy can introduce serious artifacts to the combined tight-binding model. Here, we propose two methods to seamlessly stitch two different tight-binding models that share some basis functions in common. First, we introduce a simple and efficient method: (i) finding the best matching maximally localized Wannier function pairs in the overlapping region belonging to the two tight-binding models, (ii) rotating the spin orientations of the two corresponding Wannier functions to make them parallel to each other, and (iii) making their overall phases equal. Second, we propose a more accurate and generally applicable method based on the iterative minimization of the difference between the Hamiltonian matrix elements in the overlapping region. We demonstrate our methods by applying them to the surfaces of diamond, GeTe, Bi$_2$Se$_3$, and TaAs. Our methods can be readily used to construct reliable tight-binding models for surfaces, interfaces, and defects.
\end{abstract}

\maketitle

\section{Introduction}
In the independent-particle approximation, eigenstates of a periodic electronic system
are given by 
Bloch states $\ket{\psi_{p\mathbf{k}}}$, labelled with
the band index $p$ and
the reciprocal vector index $\mathbf{k}$. Wannier functions (WFs) are
an alternative representation of the Bloch states
that provide an atomic-orbital-like,
localized basis set.
The WFs are obtained by first
performing a unitary transformation to generate $N_{\rm W}$ Bloch states at each k point
\begin{equation} \label{eq:wfdef1}
    \ket{\tilde{\psi}_{n\mathbf{k}}}
    = \sum_{p}\ket{\psi_{p\mathbf{k}}} U_{pn}(\mathbf{k})
\end{equation}
and then Fourier transforming the periodic Bloch states into localized WFs:
\begin{equation} \label{eq:wfdef2}
    \ket{w_{\mathbf{R}n}} = \frac{V_{\rm cell}}{(2\pi)^3} \int_{\rm BZ} d\mathbf{k} e^{-i\mathbf{k} \cdot \mathbf{R}} \ket{\tilde{\psi}_{n\mathbf{k}}}\,.
\end{equation}
Here, $N_{\rm W}$ is also the number of Wannier functions per unit cell,
$V_{\rm cell}$ is the volume of
the
real-space unit cell, and
the integration is performed over the Brillouin zone (BZ).
Also, the WFs are labelled by the orbital index $n$ and the unit cell index $\mathbf{R}$.
In actual calculations, the Brillouin zone is sampled with a uniform
k-point
mesh. In this case, the WFs become
\begin{equation} \label{eq:wfdef_mesh}
    \ket{w_{\mathbf{R}n}} = \frac{1}{\sqrt{N_{\rm k}}} \sum_{\mathbf k} \sum_{p}
    e^{-i\mathbf{k} \cdot \mathbf{R}} \ket{\psi_{p\mathbf{k}}} U_{pn}^{\mathbf{k}}\,,
\end{equation}
where $N_{\rm k}$ is the number of k points in the mesh.

Given a set of Bloch states, the corresponding set of WFs is not unique. This degree of freedom is parameterized by the unitary matrix $U^{\mathbf k}$ defined at each k point of the mesh. A commonly used method to determine the $U^{\mathbf k}$ matrices is to choose them such that the resulting WFs are as localized as possible in real space. The WFs that minimize the total spread functional
\begin{equation} \label{eq:spread}
    \Omega = \sum_{n=1}^{N_{\rm W}} \left[ \bra{w_{\mathbf{0}n}}r^2\ket{w_{\mathbf{0}n}} - \bra{w_{\mathbf{0}n}}\mathbf{r}\ket{w_{\mathbf{0}n}}^2 \right]
\end{equation}
are termed maximally localized Wannier functions (MLWFs)~\cite{Marzari1997MLWF,Souza2001PRB,Marzari2012RMP}.
Since the MLWFs are maximally
localized
in real space, they are often used to generate an accurate and
transferable tight-binding model of an electronic system
with the smallest possible set of hopping integrals~\cite{Marzari2012RMP}.
This tight-binding model correctly reproduces the band structure obtained by the {\it ab initio} calculation that the MLWFs are calculated from. Thus, this model is called an {\it ab initio} tight-binding model based on
MLWFs~\cite{Marzari2012RMP}.

Furthermore, one can combine the {\it ab initio} tight-binding models
based on MLWFs obtained
from different systems to construct a tight-binding model for a large-scale system. For example,
the tight-binding models for the bulk and a small supercell of a given material can be combined to describe 
disordered systems~\cite{Berlijn2011PRL},
impurities~\cite{Ku2010PRL},
interfaces~\cite{CalzolariPRB2004,Lee2005PRL,Shelley2011}, and
surfaces~\cite{Zhang2009NatPhys,Zhang2010NJP,Zhang2011PRL,Liu2013PRB}.

Among these composite systems,
we focus on the modelling of surfaces.
The electronic structure of a surface can be modelled by constructing
a tight-binding model of a finite slab or a semi-infinite
surface where the hopping parameters are obtained from
the {\it ab initio} tight-binding model based on the WFs of the {\it bulk} crystal~\cite{Zhang2010NJP,Zhang2011PRL,Liu2013PRB}.
The bulk-derived surface model correctly reproduces the topological properties
of the bulk such as the presence or absence
of topologically protected surface states. However, this model cannot take
the difference in the environment of the surface from the bulk, surface charge redistribution, and structural relaxation into account. Hence, the calculated properties of the bulk-derived surface may
significantly differ from the those of the real surface~\cite{Sun2015PRB}.

\begin{figure}
\includegraphics[width=0.95\columnwidth]{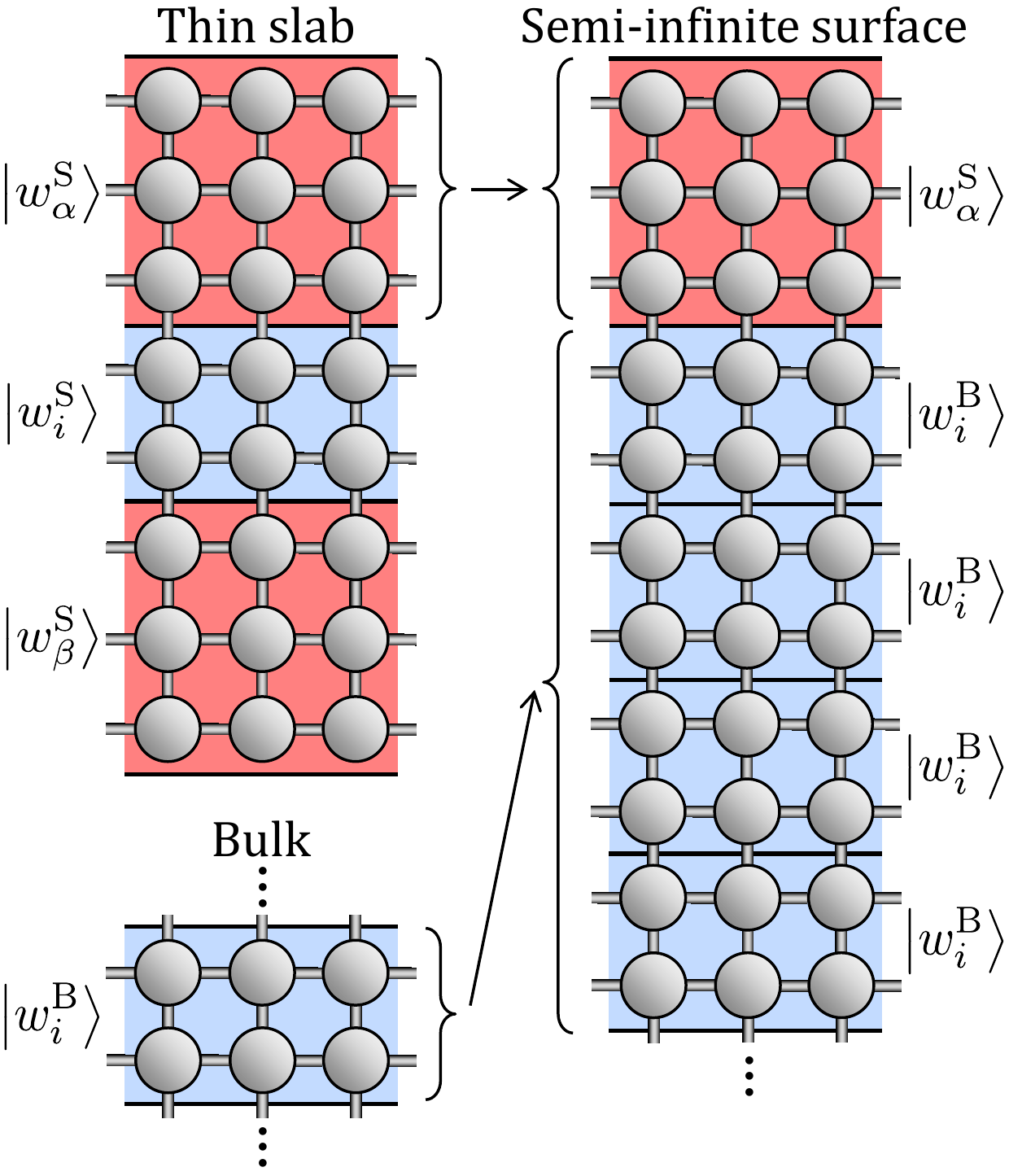}
\caption{
Schematic illustration of the construction of the tight-binding model
for a semi-infinite surface from the tight-binding models of a bulk and a thin slab.
The principal layers are indicated by shaded areas separated by
horizontal sold black lines.
The regions with red (dark gray) background correspond to the surface principal layers, in which
the surface-local perturbations to the on-site energies and
hopping parameters are present.
The regions with blue (light gray) background are bulk principal layers, where 
these surface-local perturbations are negligible.
A grey sphere represent each atomic site hosting MLWFs.
The ket vectors of the MLWFs are written alongside the principal layers where the MLWFs are located.
The superscripts
B and S indicate that the corresponding MLWFs
are the basis functions
of the tight-binding models
of the bulk and the slab, respectively.
Indices $\alpha$ and $\beta$ are used to denote the MLWFs in the surface principal 
layers,
while index $i$ is used to denote the MLWFs in the bulk principal layer.
The in-plane unit cell index $\mathbf{R}$ is omitted for clarity.
The system is assumed to be periodic in the in-plane directions
although our method does not require in-plane periodicity.
}
\label{fig:schematic}
\end{figure}

A more accurate method to describe surfaces is
to construct a tight-binding model
of a semi-infinite surface by combining the MLWF-based tight-binding models for the bulk and
a thin slab of a given material~\cite{Zhang2009NatPhys} as illustrated in Fig.~\ref{fig:schematic}.
This construction is done by dividing the system into principal layers~\cite{Lee1981PRB}.
A principal layer is a group of atomic layers which is sufficiently thick such
that the hopping between different principal layers is negligible except between the nearest neighbors.
To combine the MLWF-based tight-binding models, first, we regard the principal layer at the center of
the thin slab as a bulk principal layer.
Note that the {\it thin} slab should be {\it thick} 
enough such that the hopping
integrals in the principal layer in the middle of the slab
converge to those
in the corresponding principal layer in the true bulk system.
Then, the bottom-surface principal layer is removed
so that the slab has only the top-surface
principal layer
and one bulk principal layer.
Now, at the bottom of this composite system, we append
an infinite number of bulk principal layers to make a semi-infinite surface.
The constructed tight-binding model for the semi-infinite
surface can take the surface-local changes of the hopping parameters into account.
This scheme can be straightforwardly modified to model a thick
but finite slab by inserting
a finite number of bulk principal layers between 
the top- and bottom-surface principal
layers of the thin slab.
The electronic structure of the combined
semi-infinite surface and the finite slab
can be calculated
by iterative calculation of the surface Green function~\cite{Sancho1985} and by direct diagonalization, respectively.

When the tight-binding models of the bulk and slab based on MLWFs are combined,
the difference between the MLWFs of the two systems
should be taken care of.
Due to the maximal localization procedure, the MLWFs in the bulk and
those in the center of the slab
could be different
from each other even if the local atomic structure and the initial guess functions are the same.
Several factors can contribute to this discrepancy. First, MLWFs can be arbitrarily permuted
among themselves and be multiplied by overall phases
because these operations do not change the total spread.
Also, for calculations with spin-orbit coupling (SOC),
the corresponding MLWFs of the bulk and the slab can have different spin orientations.
Finally, the orbital part of the MLWFs obtained from the bulk and the slab can have nontrivial differences.
If these discrepancies are not properly handled, they can act as non-physical impurities at the interface where the bulk and the slab tight-binding models are combined. These impurities
can lead to artifacts in the calculated physical quantities. Therefore, it is important to correct the differences between the MLWFs before combining the tight-binding models.
Throughout this paper, we call that the bulk and slab tight-binding models based on MLWFs are ``seamlessly stitched'' if the difference between the corresponding bulk and slab MLWFs is made small.

In this paper, we propose two post-processing methods
to achieve seamless stitching of tight-binding models based on MLWFs.
The first method is a simple and efficient correction:
(i) find the best matching MLWF pairs in the overlapping principal layer of the bulk and slab tight-binding models,
(ii) rotate the spin axes of the slab MLWFs to make them parallel to those of the corresponding bulk MLWFs,
and (iii) make their overall phases equal.
The second method is based on the minimization of the difference of the on-site and hopping parameters between the bulk and the slab tight-binding models
and is more accurate and generally applicable.

The remainder of this paper is organized as follows.
In Sec.~II, we describe the methods proposed in this paper in detail.
We then illustrate the utility of our
methods by applying them to
the surfaces of diamond, GeTe, Bi$_2$Se$_3$, and TaAs.
We specify the computational details in Sec.~III, and present and discuss the results in Sec.~IV.
Another method often used to combine {\it ab initio} tight-binding models is to use the WFs
obtained without maximal localization as the basis functions \cite{Zhang2010NJP,Ku2010PRL,Berlijn2011PRL}.
These ``projection-only'' WFs are obtained by projecting the
atomic-orbital-like initial guesses for the WFs
onto the target subspace of Kohn-Sham eigenstates.
We compare the projection-only
WF method with the corrections proposed in this paper in Sec.~V.
Finally, we present a summary and conclude in Sec.~VI.

\section{Methods}
In this section, we explain our correction methods in detail.
We focus on the seamless stitching of the bulk and the slab tight-binding models
for constructing a tight-binding model of the surface.
Hence, we describe the corrections that minimize the difference between the MLWFs at the interior of the slab and the corresponding MLWFs of the bulk system of the same material.
However, the methods described in this section can be
straightforwardly extended to
other problems such as modelling of interfaces or impurities.

\subsection{Correction on the Wannier functions}
Let $\ket{w_{\mathbf{R}i}^{\rm B}}$ and $\ket{w_{\mathbf{R}n}^{\rm S}}$ denote the bulk and slab MLWFs, respectively.
Here, the index $\mathbf{R}$ denotes the bulk and slab lattice vectors
{\it along the in-plane direction},
which is orthogonal to the surface normal direction of the slab.
Indices $i \in \{1,2,...N_{\rm W}^{\rm B}\}$ and $n \in \{ 1,2,...N_{\rm W}^{\rm S}\}$,
are the orbital indices,
where $N_{\rm W}^{\rm B}$ is the number of MLWFs per in-plane unit cell of the bulk principal layer and $N_{\rm W}^{\rm S}$ is the number of MLWFs in the entire slab per in-plane unit cell.
Since the slab is thicker than the bulk principal layer along
the out-of-plane direction of the slab,
there are more MLWFs per unit cell in the slab than in the bulk,
i.\,e.\,, $N_{\rm W}^{\rm S}>N_{\rm W}^{\rm B}$.

Within this setup, our task is to find the optimal linear transformation
on $\ket{w_{\mathbf{R}n}^{\rm S}}$'s to make
new slab WFs, which we denote as $\ket{\widetilde{w}_{\mathbf{R}n}^{\rm S}}$, be
as close as possible to $\ket{w_{\mathbf{R}i}^{\rm B}}$ for $n=i$.
In other words, we want to make
\begin{equation} \label{eq:wtilde}
    \ket{\widetilde{w}_{\mathbf{R}i}^{\rm S}} \approx \ket{w_{\mathbf{R}i}^{\rm B}}
\end{equation}
for $1\le i \le N_{\rm W}^{\rm B}$.
Here, the new slab WFs $\ket{\widetilde{w}_{\mathbf{R}i}^{\rm S}}$ are the WFs in the principal layer at the center of the slab. The similarity between the bulk WFs and the slab WFs in the center of the slab directly results in the seamless stitching of the bulk and the slab WFs because the WFs in the top and bottom surfaces of the slab are already seamlessly stitched with the WFs in the center of the slab by the construction of WFs.

In general, a
linear transformation between the original and the new slab WFs is parameterized by a single $N_{\rm W}^{\rm S} N_{\rm k}^{\rm S} \times N_{\rm W}^{\rm S}$ matrix
$V$:
\begin{equation} \label{eq:vmat_def}
    \ket{\widetilde{w}_{\mathbf{R'}n}^{\rm S}}
    = \sum_{\mathbf{R}m}  \ket{w_{\mathbf{R+R'}m}^{\rm S}} V_{\mathbf{R}m,\mathbf{0}n}\,,
\end{equation}
where we have used
$V_{\mathbf{R+R'}m,\mathbf{R'}n} = V_{\mathbf{R}m,\mathbf{0}n}$.
Here, $N_{\rm k}^{\rm S}$ is the number of k points in the Monkhorst-Pack mesh of the slab.
Using Fourier transformation, one can decompose $V$ into a sum of $N_{\rm W}^{\rm S}$ dimensional square matrices defined at each k point:
\begin{equation}
V_{\mathbf{R}m,\mathbf{0}n} 
= \frac{1}{N_{\rm k}^{\rm S}} \sum_{\mathbf{k}} e^{i\mathbf{k}\cdot\mathbf{R}} V^{\mathbf{k}}_{mn}\,,
\label{eq:vmat_k1}
\end{equation}
or, equivalently,
\begin{equation}
V^{\mathbf{k}}_{mn} 
= \sum_{\mathbf{R}} e^{-i\mathbf{k}\cdot\mathbf{R}} V_{\mathbf{R}m,\mathbf{0}n}\,.
\label{eq:vmat_k2}
\end{equation}

Since the constructed slab WFs must
also form an orthonormal set, namely
\begin{equation} \label{eq:wtilde_ortho}
    \braket{\widetilde{w}_{\mathbf{R}m}^{\rm S}}{\widetilde{w}_{\mathbf{0}n}^{\rm S}} = \delta_{\mathbf{R}m,\mathbf{0}n}\,,
\end{equation}
$V$ should satisfy the constraint
\begin{equation} \label{eq:vmat_rconst}
    \sum_{\mathbf{R'},m'} V^{\dagger}_{\mathbf{R}m
    ,\mathbf{R'}m'}V_{\mathbf{R'}m',\mathbf{0}n} = \delta_{\mathbf{R}m,\mathbf{0}n}\,.
\end{equation}
This is equivalent to the unitarity of each $V^{\mathbf{k}}$ matrix
\begin{equation} \label{eq:vmat_kconst}
    V^{\mathbf{k}} V^{\mathbf{k}\dagger} = V^{\mathbf{k}\dagger} V^{\mathbf{k}} = I_{\rm S}\,,
\end{equation}
where $I_{\rm S}$ is
the $N_{\rm W}^{\rm S}$ dimensional square identity matrix.
We note that the $V$ matrix transforms the WFs within the Wannier subspace without altering the subspace itself.

Now, we define two criteria to evaluate the quality of the constructed WFs.
The first criterion is based on the orthonormality of the WFs.
If all of the new slab WFs are almost identical to the bulk MLWFs with the same orbital index, the orthonormality relation
\begin{equation} \label{eq:w_orthoapprox}
    \braket{\widetilde{w}_{\mathbf{R}n}^{\rm S}}{w_{\mathbf{0}i}^{\rm B}} 
    \approx \delta_{n,i} \delta_{\mathbf{R},\mathbf{0}}
\end{equation}
would be satisfied; hence, we may use the deviation from Eq.~\eqref{eq:w_orthoapprox}
in measuring
the resemblance between the two sets of WFs. 
We call the absolute deviation of the overlap between
the bulk WF, $\left|w_{\mathbf{0}i}^{\rm B}\right>$,
and the surface WF, $\left|\tilde{w}_{\mathbf{R}n}^{\rm S}\right>$,
from orthonormality as the ``individual overlap error'' $\left|\Delta S_{\mathbf{R}n,\mathbf{0}i}\right|$:
\begin{equation}\label{eq:overlap_inddef}
    \left| \Delta S_{\mathbf{R}n,\mathbf{0}i}  \right|
    = \left| \braket{\widetilde{w}_{\mathbf{R}n}^{\rm S}}{w_{\mathbf{0}i}^{\rm B}} - \delta_{n,i} \delta_{\mathbf{R},\mathbf{0}} \right|  \,.
\end{equation}
We also define the ``average overlap error'' $\Delta S_{\rm ave}$:
\begin{equation} \label{eq:overlap_avedef}
    \Delta S_{\rm ave}
    = \sqrt{\frac{1}{N_{\rm W}^{\rm B}}
    \sum_{i=1}^{N_{\rm W}^{\rm B}} \left(
    \sum_{\mathbf R} \sum_{n=1}^{N_{\rm W}^{\rm S}}
    \left| \Delta S_{\mathbf{R}n,\mathbf{0}i} \right| ^2\right)}\,.
\end{equation}
Note that $\Delta S_{\rm ave}$ is bounded above by 2
in case
$\braket{\widetilde{w}_{\mathbf{R}n}^{\rm S}}{w_{\mathbf{0}i}^{\rm B}}
= - \delta_{n,i} \delta_{\mathbf{R},\mathbf{0}}$
for all ${\mathbf R}$, $n$, and $i$.

To derive the relation between the overlap error and the $V$ matrix, we first define the overlap matrix $A$ as the inner product of the bulk and slab MLWFs:
\begin{equation}
A_{\mathbf{R}m,\mathbf{0}i} =
\braket{w_{\mathbf{R}m}^{\rm S}}{w_{\mathbf{0}i}^{\rm B}}\,.
\label{eq:overlap_amat1}
\end{equation}
Its Fourier transformation is
\begin{equation}
A^{\mathbf{k}}_{mi} = \sum_{\mathbf{R}} e^{-i\mathbf{k}\cdot\mathbf{R}} A_{\mathbf{R}m,\mathbf{0}i}\,,
\label{eq:overlap_amat2}
\end{equation}
and, equivalently,
\begin{equation}
A_{\mathbf{R}m,\mathbf{0}i} = \frac{1}{N_{\bf k}^{\rm S}}\sum_{\mathbf{k}} e^{i\mathbf{k}\cdot\mathbf{R}}
A^{\mathbf{k}}_{mi}\,.
\label{eq:overlap_amat3}
\end{equation}

The detailed procedures for obtaining $A$
using first-principles density-functional theory (DFT)
methods and plane-wave basis sets
are described in Appendix A.

Given the $A$ matrix, it is straightforward to calculate the overlap
error therefrom.
Using
Eqs.~\eqref{eq:vmat_def}, \eqref{eq:vmat_k1}, \eqref{eq:overlap_amat1},
and~\eqref{eq:overlap_amat3}, or
the convolution property of Fourier transformation,
one obtains
\begin{eqnarray}    \label{eq:overlap_convol}
    \braket{\widetilde{w}_{\mathbf{R}n}^{\rm S}}{w_{\mathbf{0}i}^{\rm B}} 
    &=& \sum_{\mathbf{R'},m} V^\dagger_{\mathbf{R}n,\mathbf{R'}m}\braket{w_{\mathbf{R'}m}^{\rm S}}{w_{\mathbf{0}i}^{\rm B}} \nonumber \\
    &=& \frac{1}{N_{\rm k}^{\rm S}} \sum_{\mathbf k} e^{i\mathbf{k}\cdot\mathbf{R}} \left(V^{\mathbf{k}\dagger} A^{\mathbf k}\right)_{ni}.
\end{eqnarray}
Then, the individual overlap error is
\begin{eqnarray}
    \left| \Delta S_{\mathbf{R}n,\mathbf{0}i} \right|
    &=& \left| \frac{1}{N_{\rm k}^{\rm S}} \sum_{\mathbf k} e^{i\mathbf{k}\cdot\mathbf{R}} \left(V^{\mathbf{k}\dagger} A^{\mathbf k}\right)_{ni} - \delta_{n,i} \delta_{\mathbf{R},\mathbf{0}} \right| \nonumber \\
    &=& \frac{1}{N_{\rm k}^{\rm S}} \left| \sum_{\mathbf k} e^{i\mathbf{k}\cdot\mathbf{R}} \left(V^{\mathbf{k}\dagger} A^{\mathbf k} - I_{\rm SB}\right)_{ni}  \right|,
\label{eq:overlap_ind}
\end{eqnarray}
where $(I_{\rm SB})_{ni} = \delta_{n,i}$.
A simple expression for the average overlap error is obtained using Parseval's
theorem:
\begin{equation}
    \Delta S_{\rm ave} = \sqrt{\frac{1}{N_{\rm W}^{\rm B}N_{\rm k}^{\rm S}}\sum_{\mathbf k} \fnorm{V^{\mathbf{k} \dagger}A^{\mathbf k}-I_{\rm SB}}^2}\,.
\label{eq:overlap_ave}
\end{equation}
Here, $\fnorm{X} = \sqrt{\Tr \left( XX^\dagger \right)}$ is the Frobenius norm.

The overlap error is a direct measure of the quality of the corrected WFs.
However, calculation of the overlap error can be troublesome, mainly because of the need to calculate the overlap matrix $A$. The matrix elements of $A$ are the inner products between the bulk and the slab MLWFs.
These values are not calculated in usual Wannierization procedures where the bulk and the slab MLWFs are computed separately.
In addition, in plane-wave DFT calculations,
the real-space values of the WFs are calculated
on a discrete grid.
The spacing of this grid
is determined by the lattice constants and the wavefunction energy cutoff.
If the out-of-plane lattice constants
of the bulk and the slab are incommensurate, the real-space
grids also become incommensurate.
It follows that the WFs need to be interpolated to calculate the overlap matrix $A$.

Moreover, when ultrasoft pseudopotential (USPP) or projector augmented wave (PAW) potential is used for the DFT calculation, the simple orthogonality
condition between the bulk and the slab WFs [Eq.~\eqref{eq:w_orthoapprox}]
does not hold
because only a generalized orthogonality relation between the pseudo wavefunctions is satisfied in such calculations~\cite{Vanderbilt1990PRB,Blochl1994PRB,Martin2004}.
The generalized orthogonality relation is determined from the projector functions
of the USPP or PAW potentials.
In usual electronic structure calculations, the inner product between
pseudo wavefunctions that belong to a single system can be calculated with the help of the generalized overlap matrix,
also defined by the pseudopotential.
However, since the bulk and the slab have different atomic positions, they have different
sets of projector functions. Thus, one cannot use the generalized overlap matrix to recover orthogonality between the bulk and slab
pseudo wavefunctions.
Instead, the all-electron wavefunction should be restored,
requiring a considerable computational effort.

In this paper, we circumvent these difficulties by using norm-conserving pseudopotentials
and tuning the vacuum thickness
to make the slab supercell commensurate with the bulk unit cell. But in general, it would be difficult to assess the quality of the constructed WFs using the overlap error.

Because of this difficulty, here we define the second criterion, which makes use of the Hamiltonian matrix elements of the tight-binding model. 
To combine the bulk and the slab tight-binding models, 
one has to make sure that
the slab is
thick enough so that the charge density around the center of the slab is nearly identical to that of the bulk.
Then, the Hamiltonian at the center of the slab and the Hamiltonian of the bulk will be approximately the same.
Therefore,
the difference between the Hamiltonian matrix elements
of the bulk and the slab can be attributed solely to
the difference
in the basis functions of the tight-binding models, which are the bulk and slab WFs.
This second criterion does not require computation of quantities other than the Hamiltonian matrix elements, which are calculated in standard Wannierization procedures; hence, it
can be readily used in actual calculations.

To be concrete, we define the individual hopping error
\begin{equation} \label{eq:hop_ind}
    \left| \Delta H_{\mathbf{R}i,\mathbf{0}j} \right|
    = \left| \mel{\widetilde{w}_{\mathbf{R}i}^{\rm S}}{H_{\rm S}}
    {\widetilde{w}_{\mathbf{0}j}^{\rm S}}
    - \mel{w_{\mathbf{R}i}^{\rm B}}{H_{\rm B}}{w_{\mathbf{0}j}^{\rm B}}\right|
\end{equation}
and the average hopping error
\begin{equation} \label{eq:hop_error}
    \Delta H_{\rm ave}
    = \sqrt{\frac{1}{N_{\rm W}^{\rm B}}
    \sum_{j=1}^{N_{\rm W}^{\rm B}} \left(
    \sum_{\mathbf R} \sum_{i=1}^{N_{\rm W}^{\rm B}}
    \left| \Delta H_{\mathbf{R}i,\mathbf{0}j} \right| ^2 \right)}
\end{equation}
analogously to the definition of the individual and average overlap
errors in
Eqs.~\eqref{eq:overlap_inddef} and~\eqref{eq:overlap_avedef}.
Here, $H_{\rm S}$ and $H_{\rm B}$ are the Hamiltonian operators for the bulk and the slab, respectively.
Note that only the hopping between the WFs in the central principal layer of the slab is taken into account, i.\,e.\,, $i,~j \in \{1,2,...N_{\rm W}^{\rm B}\}$.

Before calculating the hopping errors, the reference potential of the bulk and the slab calculation need to be aligned. We use the on-site energies of the MLWFs to determine the reference potential~\cite{Corsetti2011PRB}.
The difference in
the average on-site energy
\begin{equation} \label{eq:hop_align}
    \delta H_{\rm on} = \frac{1}{N_{\rm W}^{\rm B}} \sum_{i=1}^{N_{\rm W}^{\rm B}} \left( \mel{w_{\mathbf{0}i}^{\rm S}}{H_{\rm S}}{w_{\mathbf{0}i}^{\rm S}}
    - \mel{w_{\mathbf{0}i}^{\rm B}}{H_{\rm B}}{w_{\mathbf{0}i}^{\rm B}} \right)
\end{equation}
is subtracted from 
all the on-site energies of the slab tight-binding model.
Note that the reference potential of the slab is calculated using only the MLWFs at the central principal layer.

Once the average on-site potential energy of the slab tight-binding model is aligned with
that of the bulk tight-binding model,
the individual hopping errors associated with the given $V$ matrix can be calculated
from the MLWF hopping matrix elements using the convolution property,
similarly as in Eq.~\eqref{eq:overlap_ind}:
\begin{eqnarray}
    \left| \Delta H_{\mathbf{R}i,\mathbf{0}j} \right|
    &=& \sum_{\mathbf{R'},\mathbf{R''},m,n} V^{\dagger}_{\mathbf{R}i,\mathbf{R'}m} \mel{ w_{\mathbf{R'}m}^{\rm S}}{H_{\rm S}}{w_{\mathbf{R''}n}^{\rm S}}
    V_{\mathbf{R''}n, \mathbf{0}j} \nonumber \\
    &&- \mel{w_{\mathbf{R}i}^{\rm B}}{H_{\rm B}}{w_{\mathbf{0}j}^{\rm B}} \nonumber \\
    &=& \frac{1}{N_{\rm k}^{\rm S}} \sum_{\mathbf k} e^{i\mathbf{k}\cdot\mathbf{R}} \left(V^{\mathbf{k}\dagger} H_{\rm S}^{\mathbf k}V^{\mathbf{k}} - H_{\rm B}^{\mathbf k}\right)_{ij}\,.
    \label{eq:hop_convol}
\end{eqnarray}
Here, we defined
\begin{equation}
    \left( H^{\mathbf k}_{\rm S} \right)_{mn}
    =\sum_{\mathbf R} e^{-i \mathbf{k} \cdot \mathbf{R}} \mel{w_{\mathbf{R}m}^{\rm S}} {H_{\rm S}} {w_{\mathbf{0}n}^{\rm S}}
\label{eq:hop_fourier1}
\end{equation}
and
\begin{equation}
    \left( H^{\mathbf k}_{\rm B} \right)_{ij}
    =\sum_{\mathbf R} e^{-i \mathbf{k} \cdot \mathbf{R}} \mel{w_{\mathbf{R}i}^{\rm B}} {H_{\rm B}} {w_{\mathbf{0}j}^{\rm B}}\,.
\label{eq:hop_fourier2}
\end{equation}
The sum over $\mathbf{R}$ is calculated only over the in-plane lattice vectors.

The average hopping error can be calculated using Parseval's theorem:
\begin{equation} \label{eq:hop_ave}
    \Delta H_{\rm ave}
    = \sqrt{\frac{1}{N_{\rm W}^{\rm B}N_{\rm k}^{\rm S}}
    \sum_{\mathbf k} \fnorm{ I_{\rm SB}^\dagger V^{\mathbf{k}\dagger} H_{\rm S}^{\mathbf k} V^{\mathbf{k}} I_{\rm SB} - H_{\rm B}^{\mathbf k}}^2}\,.
\end{equation}

In summary, the goal for correcting the MLWFs is to find a proper $V$ matrix that constructs a new set of slab WFs that are as similar as possible to the bulk MLWFs. The overlap error and the hopping error can be used to assess the similarity between the constructed slab WFs and the bulk MLWFs.
Hence, the overlap error and the hopping error are the measure of the degree of the ``seamlessness'' of the combined tight-binding model.

\subsection{Wavefunction correction}
Now, we introduce the methods we use to determine the $V$ matrix.
The first method is to find the optimal $V$ matrix that minimizes the average overlap error.
From Eq.~\eqref{eq:overlap_ave}, it is evident that the average overlap error can be minimized by separately optimizing each $V^{\mathbf k}$ matrix.
In this method, the overlap matrix $A^{\mathbf k}$, which contains the detailed information of the Kohn-Sham wavefunctions, needs to be calculated.
Hence, we call this method the ``wavefunction correction.''

We first rewrite each summand
in Eq.~\eqref{eq:overlap_ave} as
\begin{equation}\label{eq:wfcorr_ia}
    \fnorm{V^{\mathbf{k}\dagger} A^{\mathbf k} - I_{\rm SB}}^2
    = \fnorm{A^{\mathbf k} - V^{\mathbf k} I_{\rm SB}}^2\,,
\end{equation}
using the unitarity of $V^{\mathbf k}$.
Henceforth, we temporarily omit the superscript ${\mathbf k}$ for brevity.
The above expression is minimized when
\begin{equation} \label{eq:wfcorr_constr}
    V I_{\rm SB} = U 
    \begin{bmatrix} I_{\rm B} \\ 0  \end{bmatrix}
    W^{\dagger}
\end{equation}
is satisfied~\cite{keller1975closest},
where
\begin{equation} \label{eq:wfcorr_svd}
    A = U
    \begin{bmatrix} \Sigma \\ 0  \end{bmatrix} 
    W^{\dagger}
\end{equation}
is the singular value decomposition (SVD) of $A$.
Here, $U$ and $W$ are $N_{\rm W}^{\rm S}$ and $N_{\rm W}^{\rm B}$ dimensional square unitary matrices, respectively,
and $\Sigma$ is an
$N_{\rm W}^{\rm B}$ dimensional square diagonal matrix with the singular values of $A$ on its diagonal. Also, we define $I_{\rm B}$ as the $N_{\rm W}^{\rm B}$ dimensional identity matrix.

Since $I_{\rm SB}$ is a rectangular matrix that does not have a right inverse,
$V$ is not uniquely determined from Eq.~\eqref{eq:wfcorr_constr}.
To settle the remaining degree of freedom, 
we use the fact that
the underdetermined degree of freedom can be parameterized by a unitary matrix $X$ of the form
\begin{equation}
X=\begin{bmatrix} I_{\rm B} & 0 \\ 0 & \tilde{X}
\end{bmatrix}\,,
    \label{eq:wfcorr_x}
\end{equation}
where $\tilde{X}$ is a unitary matrix of dimension
$N_{\rm W}^{\rm S}-N_{\rm W}^{\rm B}$.
Note that for every $V$ that satisfies Eq.~\eqref{eq:wfcorr_constr},
$VX$ also satisfies it.
This indeterminacy 
can be understood as follows.
If one determines
$N_{\rm W}^{\rm S} - N_{\rm W}^{\rm B}$ states in the slab MLWF
subspace that are orthogonal to the $N_{\rm W}^{\rm B}$ bulk MLWFs,
any unitary transformation among
them obviously does not change the average overlap error.

While the choice of $X$ does not alter the average overlap error, it can modify the WFs at the surface principal layers.
This change is problematic since the MLWFs from the top and bottom surface of the slab
can be mixed and become substantially delocalized
such that
they can `see' each other.
In other words, the WFs
living in the top principal layer can have non-zero hopping amplitude to the bottom principal layer,
which invalidates the construction process of the semi-infinite surface tight-binding model described
in Fig.~\ref{fig:schematic}.
To avoid this possibility, we choose $X$ so that the constructed WFs deviate as little as possible from their corresponding original MLWFs. Concretely, we choose $X$ that minimizes the deviation of $VX$ from the identity,
$\fnorm{VX - I_{\rm S}}$.

Now, we define the projection matrices $P=I_{\rm SB}I_{\rm SB}^\dagger$ and $Q=I_{\rm S} -P$. It is straightforward that $PI_{\rm SB} = I_{\rm SB}$ and $QI_{\rm SB} = 0$ hold.
Using the fact that 
\begin{equation} \label{eq:wfcorr_fnormid1}
    \fnorm{Z}^2 = \fnorm{ZP}^2 + \fnorm{ZQ}^2
\end{equation}
and
\begin{equation} \label{eq:wfcorr_fnormid2}
    \fnorm{Z}^2 = \fnorm{PZ}^2 + \fnorm{QZ}^2
\end{equation}
holds for every matrix $Z$, one can show that
\begin{eqnarray}
    &&\fnorm{VX - I_{\rm S}}^2 \nonumber \\
    &=& \fnorm{VP - P}^2 + \fnorm{VXQ - Q}^2 \nonumber \\
    &=& \fnorm{VP - P}^2 + \fnorm{PVQ}^2 + \fnorm{QVXQ - Q}^2\,.
    \label{eq:wfcorr_vxnorm}
\end{eqnarray}
Since the first two terms in the third line of Eq.~\eqref{eq:wfcorr_vxnorm} is independent of $X$,
it suffices to minimize the last term only.
Using the property $QVXQ=QVQXQ$, one can obtain the
unitary matrix $X$ that minimizes Eq.~\eqref{eq:wfcorr_vxnorm}
from the SVD of $QVQ$ inside the $Q$ subspace~\cite{horn1990matrix}.
Let's denote
\begin{equation}
QVQ = \begin{bmatrix} 0_{\rm B} & 0 \\ 0 & \tilde{V}
\end{bmatrix}\,,
    \label{eq:QVQ}
\end{equation}
where $0_{\rm B}$ is the $N_{\rm W}^{\rm B}$-dimensional null matrix.
With the SVD
\begin{equation} \label{eq:wfcorr_qvq}
    \tilde{V} = U' \Sigma' W'^{\dagger}\,,
\end{equation}
the optimal $\tilde{X}$ [Eq.~\eqref{eq:wfcorr_x}] satisfies
\begin{equation} \label{eq:wfcorr_qxq}
    \tilde{X} = W'U'^\dagger\,.
\end{equation}
This fully determines
$X$ [Eq.~\eqref{eq:wfcorr_x}] and hence $VX$.
We call this $VX$ the optimal $V$.

In summary, in the wavefunction correction method, one finds the optimal $V$ matrix that minimizes the average overlap error $\Delta S_{\rm ave}$.
This is done by first finding a unitary matrix $V^{\mathbf k}$ that satisfies Eq.~\eqref{eq:wfcorr_constr} using $U$ and $W^\dagger$ as defined in Eq.~\eqref{eq:wfcorr_svd},
and then fixing the remaining indeterminacy using  
Eqs.~\eqref{eq:wfcorr_x}, \eqref{eq:wfcorr_qvq}, and~\eqref{eq:wfcorr_qxq}.
This procedure is repeated for each $V^{\mathbf k}$.

Despite its optimality, the wavefunction correction may not be easy to use in practice; the calculation of the overlap matrix $A$ requires additional computational expense and can be problematic depending on the structure of the slab supercell and the type of the pseudopotentials, as described in Sec.~II.A.
Since we aim at post-processing methods that work irrespective of the computational details,
we use the wavefunction correction
only as a reference method to evaluate other methods.

\subsection{Minimal correction}
The first practical approach we propose is to correct only the easily tractable degrees of freedom of the MLWFs, which are the permutation order,
the spin axis, and the overall phase.
We call this method  ``minimal correction'' because the applied operations 
require little computational cost and
are the necessary minimum to obtain a reliable tight-binding 
model, as will be demonstrated with real material examples in Sec.~IV.A.
In this method, it is assumed that the corresponding MLWFs of the bulk and the slab have
almost the same orbital 
wavefunctions up to an overall phase.
This assumption justifies the minimal correction in that only the permutation, spin, and overall phase degrees of freedom remain to be corrected.

Now, we describe the procedures in detail.
First, for calculations with SOC, the minimal correction consists of three steps:
pairing, spin correction, and phase correction.
In the pairing step,
the permutation degree of freedom is fixed
by finding the best matching bulk and slab MLWF pairs.
The MLWFs having the same orbital parts and orthogonal spinor parts
are not distinguished in this step.
The matching between the bulk and the slab MLWFs can be done by
pairing the MLWFs with similar center positions.

If multiple MLWFs (other than
the ones with
the same orbital parts but orthogonal spinor parts) have similar centers, the ``signatures''~\cite{Shelley2011} of the MLWFs are exploited to find the pairs.
The signatures of an MLWF are the
Fourier components
of the MLWF in the plane-wave representation.
Concretely, the signature $I_{m}(\mathbf{G})$ of
an
MLWF $\ket{w_{\mathbf{0}m}}$ for a given reciprocal lattice vector $\mathbf{G}$ is
\begin{equation} \label{eq:mincorr_sig}
    I_{m}(\mathbf{G}) = \frac{1}{\sqrt{V_{\rm cell}}} \int_{V_{\rm cell}} d\mathbf{r} e^{-i \mathbf{G}\cdot(\mathbf{r}-\mathbf{r_{\rm c}})} \braket{\mathbf r}{w_{\mathbf{0}m}}\,.
\end{equation}
Here, ${\mathbf r_{\rm c}}$ is the center of the MLWF $\ket{w_{\mathbf{0}m}}$.
In practice, the signatures of the MLWFs are calculated for a few
{\bf G} vectors close to $\mathbf{0}$.
We calculated WF signatures at five $\mathbf{G}$'s: (0,0,0), ($\pm$1,0,0), and (0,$\pm$1,0) in the reciprocal lattice coordinate.
If necessary, one can add more {\bf G} vectors to the list.
The method we use to calculate the MLWF signatures is detailed in Appendix A.

Since the signatures of an
MLWF encode its orbital character, one can use the signatures to pair the MLWFs.
For example, assume that a p$_x$-orbital-like MLWF and a p$_y$-orbital-like MLWF
are centered at the same atom.
The two WFs can be distinguished using the following signature
\begin{eqnarray}
    &&I_{m}(\mathbf{G}_x)-I_{m}(-\mathbf{G}_x) \nonumber \\
    &&= \frac{-2i}{\sqrt{V_{\rm cell}}} \int_{V_{\rm cell}} d\mathbf{r}
    \sin (\frac{x-x_{\rm c}}{L_x})
    \braket{\mathbf r}{w_{\mathbf{0}m}}\,,
     \label{eq:mincorr_psig}
\end{eqnarray}
where $\mathbf{G}_x = 2\pi / L_x \hat{x}$.
Obviously, the absolute value of this quantity
of a p$_x$-like MLWF will be much larger than that of a p$_y$-like MLWF.

When multiple MLWFs have the same center, we manually 
pair the bulk and slab MLWFs with similar signatures.
The validity of the constructed pairs can be judged by checking whether the average hopping error after the minimal correction is reasonably small.

The next step is the spin correction. In this step, the spin quantization axes of the bulk and the slab MLWFs are aligned with each other.
Also, the relative phases between the spin up and spin down MLWFs of the bulk and the slab are made equal.
This is done by calculating the matrix elements of the spin operator in the MLWF basis.
For MLWFs $\ket{w_+}$ and $\ket{w_-}$
that have almost the same orbital wavefunction but mutually orthogonal spinors,
let us write the matrix elements of the spin angular momentum operator $\mathbf{S}$ as
\begin{equation}
    \mathbf{S}_{\sigma,\sigma'} = \mel{w_\sigma}{\mathbf{S}}{w_{\sigma'}}\,,
\end{equation}
where $\sigma, \sigma' \in \{+,-\}$.
The diagonal matrix elements, $\mathbf{S}_{+,+}$ and $\mathbf{S}_{-,-}$,
are the expectation values of the spin polarization 
of the MLWFs.
Especially, the orientation
of this expectation value is of our interest.
The relative phase between $\ket{w_+}$ and $\ket{w_-}$ 
can be calculated from the off-diagonal matrix elements
$\mathbf{S}_{+,-}$.
We note that the spin matrix elements can be calculated using the Wannier90 package~\cite{mostofi2014updated}.
After the two, opposite spin orientations and the relative phase for each such $\ket{w_+}$ and $\ket{w_-}$ of the bulk and the slab MLWFs are calculated, one can multiply an appropriate block-diagonal
unitary matrix composed of $2\times2$ spin-rotational unitary matrices
to the $V$ matrix so that the spinor parts of the bulk and 
the corresponding slab MLWFs become the same.
However, the two states
in the slab may still have a different common overall phase than the corresponding two
states in the bulk.

The last step of the minimal correction is the phase correction.
Here, the common overall phase of the MLWFs
$\ket{w_+}$ and $\ket{w_-}$ is calculated from
their signatures.
Consider two bulk MLWFs $\ket{w_+}$ and $\ket{w_-}$
and the corresponding
two slab MLWFs which are paired with those
bulk MLWFs in the pairing step.
After the spin correction step, these bulk and slab MLWFs differ only by a single overall phase.
Now, let $I_{\sigma}^{\rm B}(\mathbf G)$ and $I_{\sigma}^{\rm S}(\mathbf G)$ be the signatures of the bulk and slab MLWFs, respectively, where $\sigma \in \{+,-\}$ is the spinor index. 
Then, the difference of the the overall phase between the bulk and the slab MLWFs can be approximated as
\begin{equation}
    \phi = \arg \sum_{\mathbf{G},\sigma}
I_{\sigma}^{\rm B}(\mathbf G) I_{\sigma}^{\rm{S}}(\mathbf G)^*\,,
\label{eq:phi}
\end{equation}
where $\arg x$ is the argument of a complex number $x$.
Note that the sum runs only over the five
$\mathbf{G}$ vectors for which
the MLWF signatures are calculated.
By multiplying the two slab MLWFs
by $e^{i \phi}$, one can make the overall phase of the two slab MLWFs identical to the overall phase of the two bulk MLWFs.
This finishes the correction of the permutation, spinor, and phase degrees of freedom.

For systems without SOC, the spin correction is omitted.
This also holds for systems with collinear magnetism, since the spin up and spin down WFs are completely decoupled.
If one assumes that the MLWFs are real-valued functions, one 
has to determine only the overall sign, not the overall phase
of the MLWFs.
We note that this  simplified correction
scheme based on the sign determination
is implemented in the transport module of the Wannier90 package~\cite{Shelley2011}.
However, even spinless MLWFs can in principle have a complex overall phase, so we do not skip the phase correction.

One may skip the spin correction for systems with SOC if
the spin orientations of the MLWFs remain unchanged from those of the initial guesses
(which are equally set for the pair MLWFs in the bulk and in the slab) during the process of maximal localization.
This is the case for all the materials we have tested
which are all non-magnetic.
However, this assumption may break down for materials with strong SOC or noncollinear magnetism because the MLWFs
may have preferred spin axes.
Also, one still needs to calculate the spin matrix elements to check the validity of this assumption.
Hence, we do not skip the spin correction.

The minimal correction is efficient in that it requires
no additional quantities to be calculated other than
the spin matrix elements and the MLWF signatures.
These quantities can be easily obtained in usual Wannierization procedures with little additional computational effort.
However, minimal correction is not applicable when the bulk and slab MLWFs have significantly different orbital wavefunctions,
because then the bulk and slab MLWFs cannot be paired.
Hence, it is necessary to use the same initial guesses for the bulk and the slab MLWFs.
But we also find cases where the orbital parts of the bulk and slab MLWFs significantly
differ from each other even though their initial guesses are
the same.
In these cases, the minimial correction is not applicable.

\subsection{Hamiltonian correction}
We now describe our second method, which
is more effective and generally applicable. In this method, we minimize the average hopping error $\Delta H_{\rm ave}$, and we call it ``Hamiltonian
correction.''
Compared to the wavefunction correction where the overlap error is minimized, the Hamiltonian correction is an approximate method. This is because while the overlap error is a direct measure of the nonorthogonality between the bulk and slab WFs, the hopping error is an indirect measure of the discrepancy between the WFs.
However, as long as the slab is thick enough such that the charge density
around the central principal layer of the slab is very close to the corresponding
charge density of the bulk, which is the situation we are interested in anyway,
this method is a good replacement of the wavefunction correction.
The Hamiltonian correction does not require any quantities other than the Hamiltonian matrix elements to be calculated. Therefore, the Hamiltonian correction is more realistic and efficient than the wavefunction correction.

Now, we describe the Hamiltonian correction in detail. A direct minimization of $\Delta H_{\rm ave}$ under only the unitarity constraint on $V^{\mathbf k}$ [Eq.~\eqref{eq:vmat_kconst}] is not
desirable
because the problem is badly underdetermined. To reveal this indeterminacy,
we first note that $\Delta H_{\rm ave}$ is minimized by separately minimizing each term in
Eq.~\eqref{eq:hop_ave}. Now, assume that $H^{\mathbf k}_{B}$ is a diagonal matrix.
Then, for an arbitrary diagonal unitary matrix $U$
and a square matrix $M$ of the same dimension as $U$,
$\fnorm{M} = \fnorm{U^\dagger MU}$
and $U^\dagger H^{\mathbf k}_{B} U = H_{\rm B}^{\mathbf{k}}$ hold.
It follows that $\Delta H_{\rm ave}$ is invariant to the change $V^{\mathbf k} \rightarrow V^{\mathbf k} U$ for every diagonal unitary matrix $U$. For a general case where $H^{\mathbf k}_{B}$ is a hermitian matrix diagonalized as $H^{\mathbf k}_{B} = W^\dagger D W$, the transformation $V^{\mathbf k} \rightarrow V^{\mathbf k} W^\dagger U W$ leaves $\Delta H_{\rm ave}$ unchanged and hence
represents the indeterminacy. Since $U$ can be arbitrarily chosen
at each k point,
it may make $V^{\mathbf k}$ to vary widely among the k points, and thus greatly delocalize the WFs.

To avoid this indeterminacy,
we further constrain $V$ such that $V_{\mathbf{R}m,\mathbf{0}n} \neq 0$ only if $\ket{w^S_{\mathbf{R}m}}$ and $\ket{w^S_{\mathbf{0}n}}$ are centered
on the same atom.
Assuming that all the WFs
on the same atom
reside in the same unit cell, this constraint implies
\begin{equation} \label{eq:hamcor_vconstr1}
    V_{\mathbf{R}i,\mathbf{0}j} = V^{(0)}_{ij}\delta_{\mathbf{R},\mathbf{0}}\,,
\end{equation}
and
\begin{equation} \label{eq:hamcor_vconstr2}
    V^{\mathbf{k}}_{ij} = V^{(0)}_{ij}\,,
\end{equation}
where $V^{(0)}$ is a square
block-diagonal unitary matrix. The second equation indicates that $V^{\mathbf k}$ is identical for all ${\mathbf k}$. Each block of $V^{(0)}$ corresponds to the set of WFs localized at the same atom.
This restriction
greatly reduces the number of free parameters and eliminates the risk of
delocalizing the WFs.
Due to this block-diagonality constraint, only the Wannier functions in the center of the slab is considered during the Hamiltonian correction.

We impose this block-diagonality constraint and minimize $\Delta H_{\rm ave}$ with respect to $V^{(0)}$. In this setting, $V^{(0)}$ is uniquely determined up to
a single overall phase common to all WFs. This phase does not make any difference in the calculated physical quantities such as the spectral function.

There may be cases in which the WFs are not centered on the atoms, like the bonding and anti-bonding orbitals. In such cases, the block-diagonality constraint
may be modified so that only the blocks of $V^{(0)}$ between MLWFs with similar centers are allowed to be nonzero.

We note that, by construction, the block-diagonal
unitary matrix $V^{(0)}$ cannot take
the mixing of WFs located at different positions into account.
A hypothetical problematic example for our restriction on $V^{(0)}$ will be the case of diamond in which the bulk MLWFs are atom-centered sp$^3$ orbitals, while the slab MLWFs are bonding and anti-bonding orbitals. To reconstruct bonding and anti-bonding orbitals from the atomic sp$^3$ orbitals, sp$^3$ orbitals localized at different atoms need to be linearly combined, which is forbidden in our scheme.
However, this hypothetical example of bonding and anti-bonding MLWFs of diamond is not of our concern because the bonding and anti-bonding orbitals will change into the sp$^3$ orbitals during maximal localization.

To minimize $\Delta H_{\rm ave}$ under the unitarity constraint, we use the Riemannian steepest descent method introduced in Ref.~\cite{abrudan2008steepest}. The detailed algorithm is explained in Appendix B. The initial condition for the steepest descent minimization is chosen to be the output of the minimal correction. When the minimal correction is not applicable,
the identity matrix is used as the initial guess for $V^{(0)}$.

In an actual correction procedure, the summation over
indices $i$ and $j$
in Eq.~\eqref{eq:hop_error}
may run over WFs other than the ones
in one principal layer of the bulk and the corresponding WFs
in the central principal layer of the slab.
In general, one can include additional atomic layers outside the central principal layer of the slab,
and accordingly outside the corresponding principal layer of the bulk
to compute $\Delta H_{\rm ave}$. The corresponding optimal $V^{(0)}$ includes additional block-diagonal parts for the additional atomic layers.
Although these additional parts of $V^{(0)}$ are not used in the end
for the combination of the two tight-binding models, this
extension may still improve the performance of the correction since the atoms at the boundary
and those in the middle of the central principal layer of the slab are treated on a more equal footing.
In this work, we include the additional MLWFs centered at two atomic layers both right above and right below the central principal layer.

\section{Computational Details}
To demonstrate the utility of the proposed corrections, we applied them to the surfaces of four real materials: diamond, GeTe, Bi$_2$Se$_3$, and TaAs.
To perform electronic structure calculations, we used DFT with plane-wave basis as implemented in the QUANTUM ESPRESSO package~\cite{Giannozzi2009QE}.
The exchange-correlation energy was treated within the generalized gradient approximation,
using the parameterization scheme of Perdew, Burke, and Ernzerhof~\cite{PerdewPRL1996PBE}.
The plane-wave energy cutoff for wavefunctions was set to 70 Ry for all materials.
Fully relativistic pseudopotentials for C, Ge, and Te were taken from pslibrary.0.3.1~\cite{Kucukbenli2014pslibrary031}.
Those for Ta and As were taken from the SG15 library~\cite{Scherpelz2016SG15}, and those for Bi and Se were generated using ld1.x atomic code of the QUANTUM
ESPRESSO package~\cite{Giannozzi2009QE}. 
Noncollinear spin polarization and SOC were taken into account for GeTe, Bi$\rm_2$Se$\rm_3$, and TaAs.
Magnetism was not considered.

In all calculations, we used experimental lattice constants.
The position of the atoms of the bulk structure were fully relaxed, while the structural relaxation of the surface from the bulk structure
was intentionally not taken into account
for a clear comparison of the results obtained from
different methods.
The surface we studied are (111) surface of diamond with dangling bonds, Te-terminated (111) surface of GeTe,
(0001) surface of Bi$_2$Se$_3$
which is the natural cleavage surface,
and As-terminated (001) surface of TaAs.
For diamond, GeTe, Bi$_2$Se$_3$, and TaAs, the bulk principal
layers are consisted of 6, 6, 15, and 8 atoms,
respectively,
and the slab supercells contain 42, 31, 40, and 40 atoms, respectively.
For self-consistent calculation of the electronic structures, we sampled the Brillouin zone with a uniform $12 \times 12 \times n_3$ mesh for the bulk, and $12 \times 12 \times 1$ mesh for the slab. The parameter $n_3$ was set to 11, 6, 3, and 8 for diamond, GeTe, Bi$_2$Se$_3$, and TaAs, respectively.

We used Wannier90 pacakage~\cite{mostofi2014updated} to construct the MLWFs. 
During this procedure, the Brillouin zone was sampled with
uniform $7 \times 7 \times n_3'$ and $7 \times 7 \times 1$
meshes for the bulk and the slab respectively, where $n_3'$=9, 6, 3, and 6 for diamond, GeTe, Bi$_2$Se$_3$, and TaAs, respectively.
The initial guesses for the construction of the MLWFs were atom-centered sp$^3$ orbitals for diamond, $s$ and $p$ orbitals for GeTe, $p$ orbitals for Bi$_2$Se$_3$, and Ta-centered $d$ orbitals and As-centered $p$ orbitals for TaAs.
For spin noncollinear systems, the spinor part of the initial guess orbitals were aligned along the $z$ axis, unless otherwise specified.
The inner frozen windows were set to [$-1$, 1]~eV around the Fermi level.
For insulators, the Fermi level was defined as the average of the valence band maximum and conduction band minimum energy.
The outer disentanglement windows were set to [$-\infty$, +15], [$-\infty$, +10], [$-7$, 13], and [$-9.5$, 9.5]~eV for diamond, GeTe, Bi$_2$Se$_3$, and
TaAs, respectively. 
Here, $-\infty$ indicates that the lower bound for the window was not set.
The energy windows were chosen so that the unbound states of the slab were excluded from the frozen window, and the bands that were
not used for Wannierization were excluded from the disentanglement window.

\begin{figure}
\includegraphics[width=0.95\columnwidth]{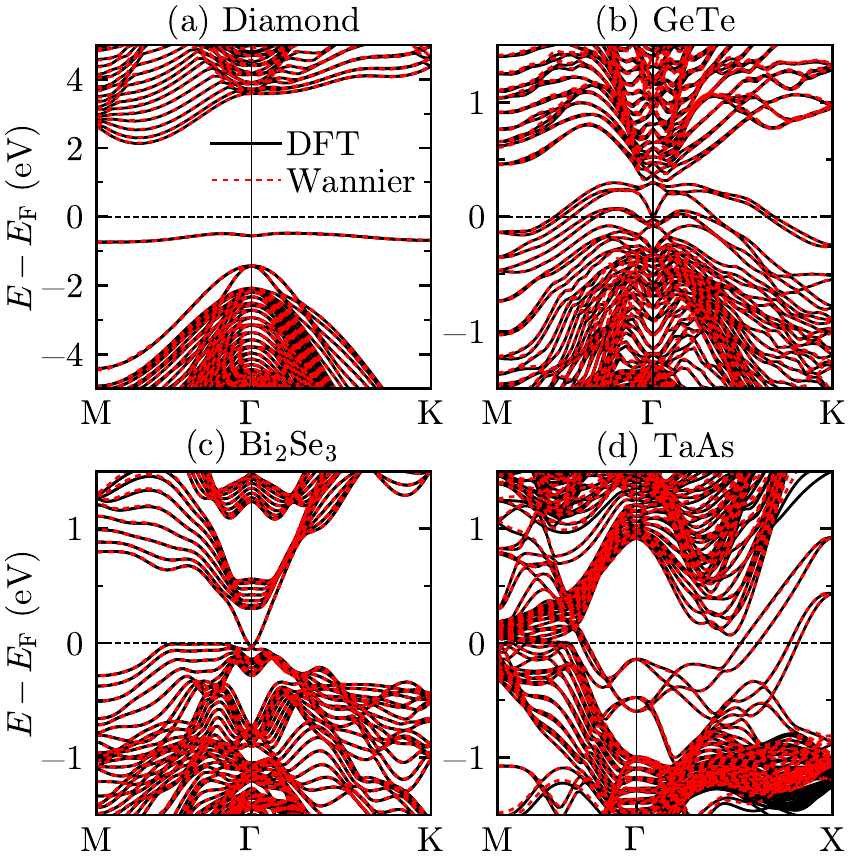}
\caption{
The band structures of the slabs obtained by DFT and by Wannier interpolation.
}
\label{fig:band}
\end{figure}

Figure~\ref{fig:band} shows that the {\it ab initio} tight-binding models of the slab based on MLWFs accurately
describe the band structure obtained from the DFT calculations inside the inner frozen energy window.
The minimal or the Hamiltonian correction to the slab tight-binding model for combination with the bulk tight-binding model is a unitary transformation common to all k points in the coarse grid.
Therefore, the energy eigenvalues are not affected by the minimal or the Hamiltonian correction.
In contrast, the wavefunction correction applies different unitary transformation to each k point in the coarse grid.
Thus, the energy eigenvalues of the tight-binding model at k points which do not belong to the coarse k-point mesh used to generate the MLWFs are also changed.
We find that for all the materials we have tested, this change in the energy eigenvalues is less than 14~meV.
Since this change is not noticeable in the scale of Fig. \ref{fig:band}, we have presented only the band structure obtained from the {\it ab initio} tight-binding models of the slab before any correction.

\section{Applications}
Now we illustrate the application of our post-processing corrections to
the surfaces of
diamond, GeTe, Bi$_2$Se$_3$, and TaAs.
Diamond is chosen to represent materials with strong covalent bonding.
For other three materials, GeTe~\cite{Liebmann2016AdvMater,Elmers2016PRB,Krempasky2018PRX}, Bi$\rm_2$Se$\rm_3$~\cite{Zhang2009NatPhys,Xia2009NatPhys}, and TaAs~\cite{Huang2015NatComm, Weng2015PRX, Xu2015Science, Lv2015PRX}, SOC have significant effects on their electronic structures, and their surface states are being actively investigated.

\subsection{Minimal correction}
\begin{figure}
\includegraphics[width=0.95\columnwidth]{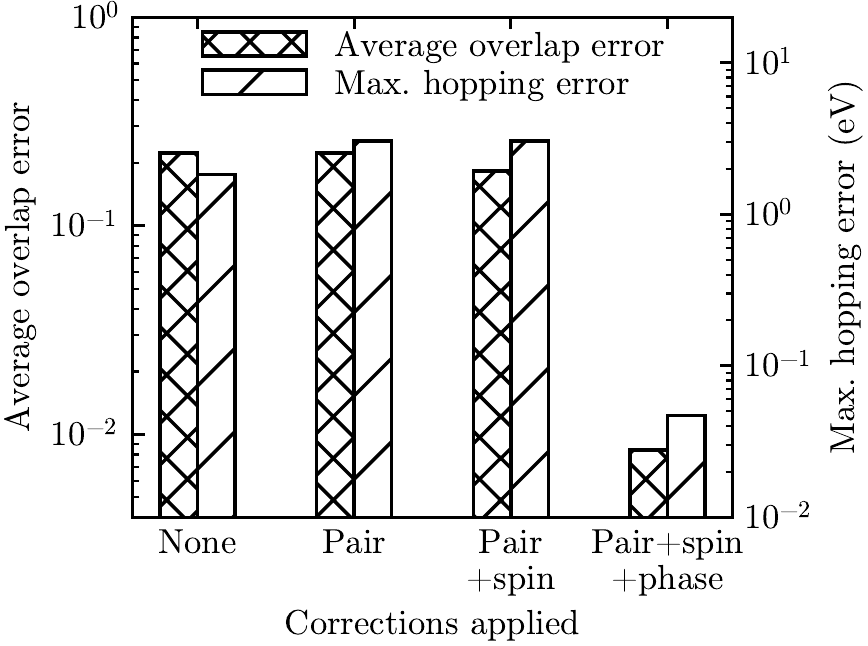}
\caption{
The overlap and hopping errors for
Bi$_2$Se$_3$ as pairing, spin correction, and phase correction are applied sequentially. The spin of the initial guesses for the bulk and slab MLWFs are aligned along the $x$
and the $z$ axes, respectively.
}
\label{fig:min_corr}
\end{figure}

We first investigate the performance of the minimal correction. In Fig.~\ref{fig:min_corr}, we show the change of the overlap and hopping errors of Bi$_2$Se$_3$ as the three steps of minimal correction, pairing, spin correction, and phase correction, are applied sequentially.
To mimic the case in which the spin axis of the MLWFs considerably differ from that of the initial guesses, for example, complex materials with strong SOC or noncollinear magnetism,
we align the spin orientations of the initial guesses of the bulk MLWFs along the $x$ axis,
and that of the slab MLWFs along the $z$ axis.

We find that the overlap and hopping errors are reduced only if all the three steps are applied. The residual maximum hopping error after the minimal correction is less than 3\% of a typical nearest neighbor hopping energy, indicating that the bulk and slab tight-binding models are properly combined. This result justifies our naming of the minimal correction, as it is the smallest set of corrections required for the seamless stitching of tight-binding models based on MLWFs.

\begin{figure}
\includegraphics[width=1.0\columnwidth]{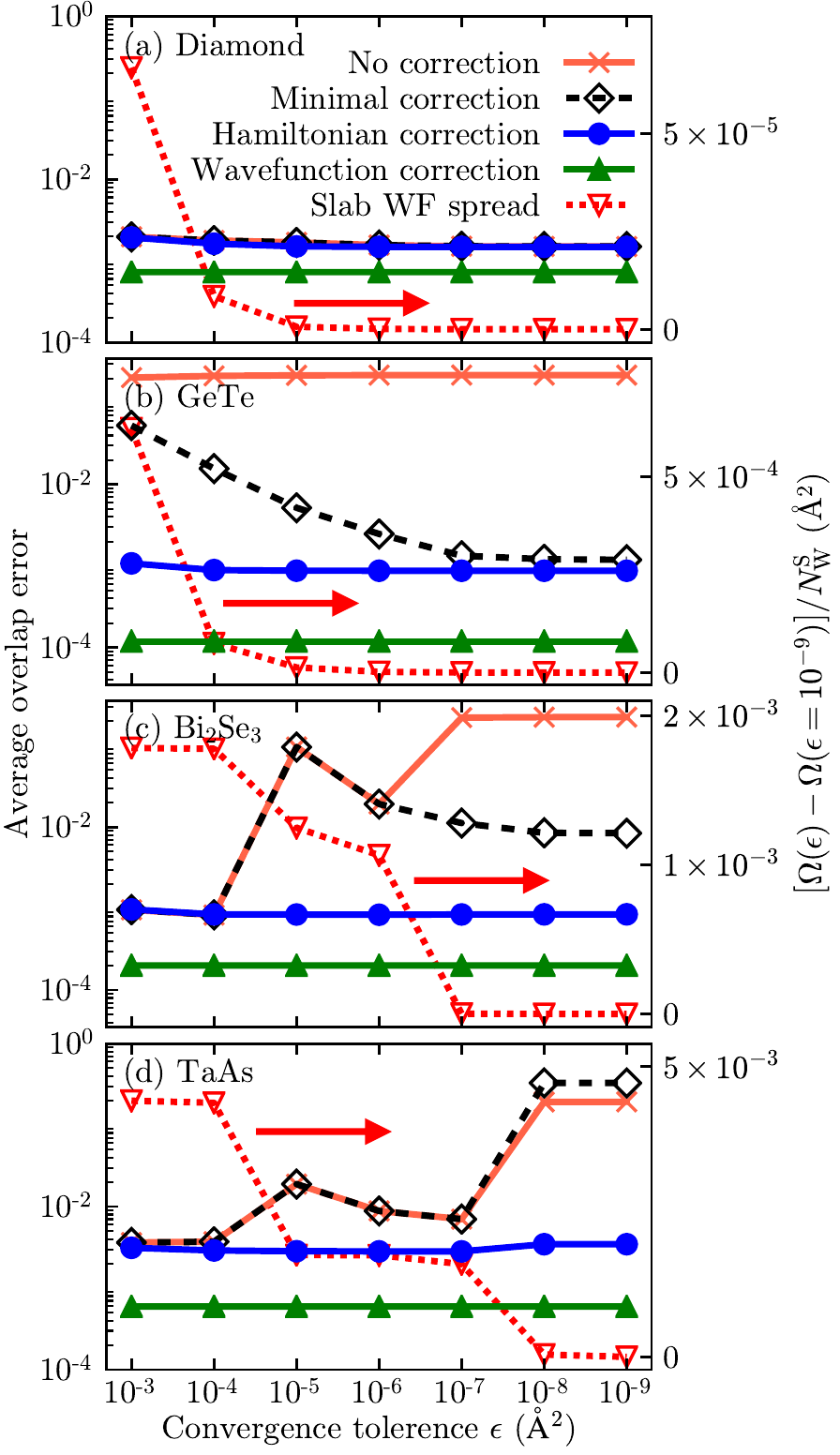}
\caption{
Dependence of the average overlap error and the WF spread of the slab upon convergence criterion, parameterized by the convergence tolerance $\epsilon$.
The total spread is normalized by dividing it with the number of MLWFs. Lines are a guide to the eye.
}
\label{fig:conv_tol}
\end{figure}

In Fig.~\ref{fig:conv_tol}, we show the dependence of the overlap error and the final WF spread $\Omega$ on the convergence criterion.
In the Wannier90 package~\cite{mostofi2014updated},
the convergence criterion is
controlled by a parameter called convergence tolerance $\epsilon$;
convergence is reached when the change in $\Omega$ per
each iteration is less than $\epsilon$ for five successive iterations.
For all the systems we have tested, the overlap and hopping errors calculated for the MLWFs obtained with $\epsilon < 10^{-9}$ do not show noticeable change in the scale shown in Fig.~\ref{fig:conv_tol} from those calculated for the MLWFs obtained with $\epsilon=10^{-9}$.
We plot the difference between $\Omega$ at given $\epsilon$ and the value of $\Omega$ at $\epsilon=10^{-9}$ for the slab MLWFs in Fig.~\ref{fig:conv_tol}. The spread $\Omega$ is normalized by dividing it by $N_{\rm W}^{\rm S}$.

For Bi$_2$Se$_3$ and TaAs, the overlap error is small even without any
correction if a sufficiently loose convergence criterion is used ($\epsilon \ge 10^{-4}$). However, this case with large $\epsilon$ is not of our interest because the constructed MLWFs are not localized
enough, as the large value of $\Omega(\epsilon)$ indicates.

We focus on the regime of sufficiently small $\epsilon$, where $\Omega$ has almost converged.
In this regime, the average overlap error of 
the MLWFs of diamond, the simplest material among the four we studied,
is almost insensitive to the convergence threshold
and is close to its minimal value obtained with the wavefunction correction. In contrast, for GeTe, the overlap error without any correction is markedly large for all values of $\epsilon$. The error is significantly reduced by the minimal correction, especially when
$\epsilon$ is small.
A similar tendency shows up in the case of Bi$_2$Se$_3$, where the
overlap error is sizable without correction but is reduced by a few orders of magnitude after the minimal correction. Also, as in the case of GeTe, the overlap error after minimal correction decreases as $\epsilon$ is lowered.
This result indicates that the orbital parts of the bulk and slab MLWFs become similar with a tight convergence criterion.
Thus, a sufficiently tight convergence criterion is required for the minimal correction to be effective.

On the contrary, in the case of TaAs, the minimal correction fails if $\epsilon \le 10^{-8}$ as the large overlap error indicates.
Upon inspection of the center position of the MLWFs, we find that for the bulk all 10 MLWFs corresponding to the Ta $d$ orbitals are precisely centered at the Ta atom. In contrast, for the slab, 4 out of 10 corresponding MLWFs are shifted from the Ta site by 0.17~\AA.
This difference signifies that the orbital parts
of the bulk and slab MLWFs that correspond to the Ta $d$
orbitals have significant differences, breaking the fundamental assumption of the minimal correction. Therefore, the minimal correction cannot be applied in this case.

In summary, the minimal correction is the smallest set of procedures required to obtain a reliable 
combined tight-binding model.
This method is simple and efficient, given that a sufficiently tight convergence criterion for localization is used. However, the minimal correction may not work when the orbital
parts of the bulk and slab MLWFs significantly differ
from each other.
Therefore, it is desirable to have a more generally useful correction method.

\subsection{Hamiltonian correction}
We now turn to the analysis of the Hamiltonian correction which is proposed to be a widely applicable and accurate. Figure~\ref{fig:conv_tol}
shows that the overlap error after Hamiltonian correction is
consistently low for all materials in the entire range of $\epsilon$.
This result demonstrates that the overlap error between MLWFs localized at the same atom gives significant contribution to the average overlap error. This error is successfully fixed by the Hamiltonian correction, while cannot be handled by the minimal correction in some cases.

We further investigate the effect of Hamiltonian correction on the overlap and the hopping errors. As we aim to compare the outcomes of the minimal correction and the Hamiltonian correction, we choose $\epsilon$ to be the smallest in so far as the minimal correction does not fail. Concretely, $\epsilon=10^{-7}$ was used for TaAs, and $\epsilon=10^{-9}$ was used for all other materials.

\begin{figure}
\includegraphics[width=0.95\columnwidth]{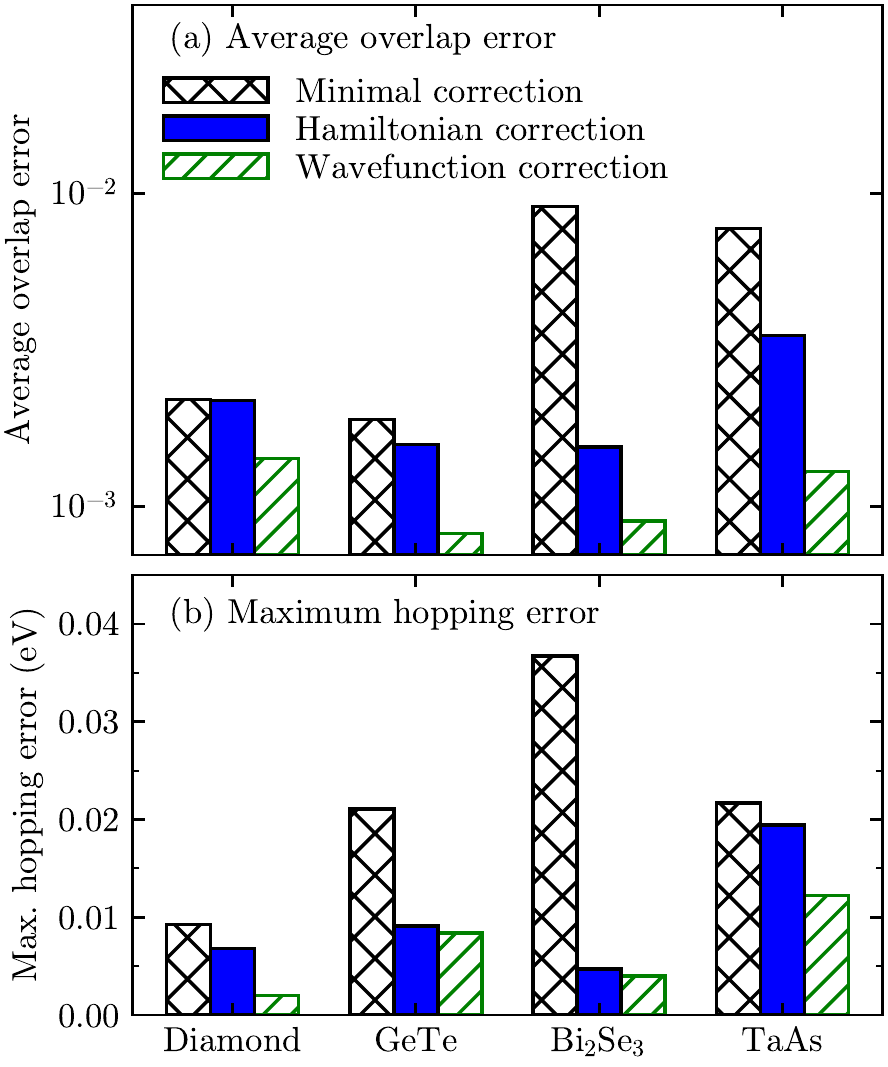}
\caption{
The (a) average overlap error and (b) maximum hopping error for different materials after either minimal, Hamiltonian, or wavefunction correction.
}
\label{fig:mlwf_error}
\end{figure}

In Fig.~\ref{fig:mlwf_error}, we show the average overlap error and the maximum hopping error for the four materials we have tested. The overlap error
after the Hamiltonian correction
is significantly smaller than the overlap error after the minimal correction. Since the Hamiltonian correction minimizes the average hopping error not the average overlap error, the decrease of the latter is
a non-trivial result, albeit expected.
The hopping error is also consistently
reduced after the Hamiltonian correction, and even becomes comparable to the reference value obtained from the wavefunction correction
in most cases.
The maximal hopping error after the Hamiltonian correction
is less than 1\% of the typical hopping
matrix element between MLWFs centered at nearest neighboring atoms.

\begin{figure}
\includegraphics[width=1.0\columnwidth]{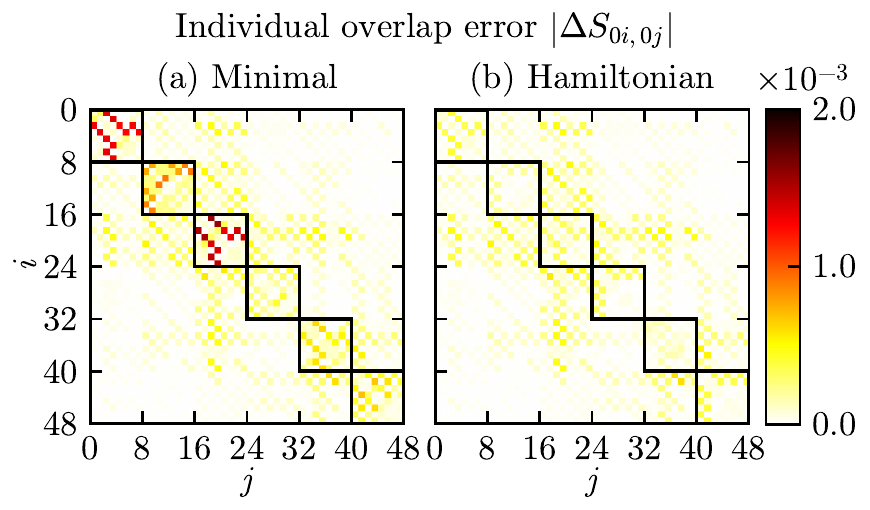}
\caption{
Individual overlap errors of GeTe as defined in Eq.~\eqref{eq:overlap_inddef} calculated between WFs in the same unit cell. The WFs are corrected using (a) the minimal correction or (b) the Hamiltonian correction. The block diagonal boxes indicate the WFs centered at the same atom.
}
\label{fig:overlap_matrix}
\end{figure}

For further analysis, we show in Fig.~\ref{fig:overlap_matrix} the individual overlap errors between MLWFs that belong to the same unit cell for GeTe. We indicate the matrix elements between MLWFs centered at the same atom using the block diagonal boxes.
While there are certain individual overlap errors relatively bigger than others after only the minimal correction, all of them are much reduced
after applying the Hamiltonian correction. This reduction in error
visually demonstrates that the Hamiltonian correction successfully corrects the dominant source of overlap error.

\begin{figure}
\includegraphics[width=1.0\columnwidth]{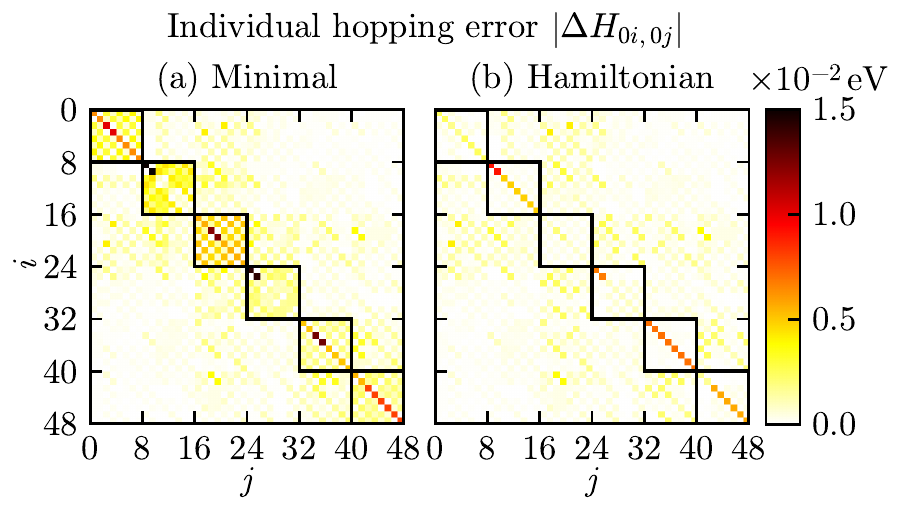}
\caption{
Individual hopping errors of GeTe as defined in Eq.~\eqref{eq:hop_ind} calculated between WFs in the same unit cell. The WFs are corrected using (a) the minimal correction or (b) the Hamiltonian correction. The block diagonal boxes indicate the WFs centered at the same atom.
}
\label{fig:hop_matrix}
\end{figure}

In Fig.~\ref{fig:hop_matrix}, we show an analogous plot for the individual hopping errors. Again, the individual hopping errors remaining after the minimal correction are significantly reduced by the Hamiltonian correction. Majority of the remaining hopping errors shown in Fig.~\ref{fig:hop_matrix}(b) are the shift of the on-site energy. We attribute this error to the effect of the surface, making the potential energy at the center of the slab differ from that of the bulk. This error could be systematically reduced if the {\it ab initio} tight-binding model is generated using a thicker slab.

\subsection{Calculation of the momentum-resolved surface local density of states}

\begin{figure*}
\includegraphics[width=0.95\textwidth]{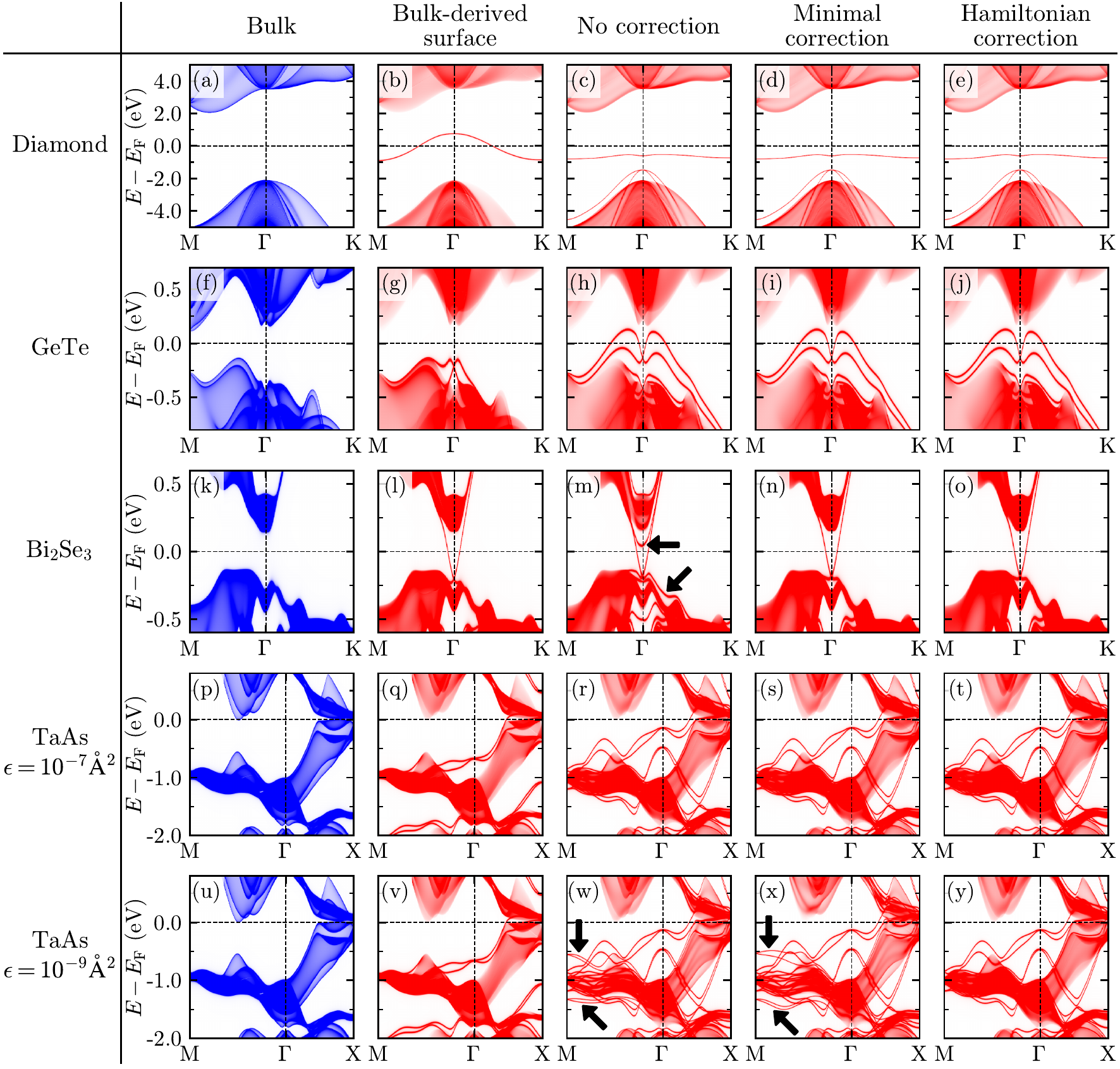}
\caption{
Momentum-resolved (a) bulk and (b)-(e) surface DOS of diamond obtained by iterative calculations of Green functions of  semi-infinite surfaces. The semi-infinite
surface is either (b) derived from the bulk tight-binding parameters, (c) derived from the slab without any correction, (d) derived from the slab with the minimal correction, or (e) derived from the slab with the Hamiltonian correction. Similar quantities
as in (a)-(e) for (f)-(j) GeTe,
(k)-(o) Bi$_2$Se$_3$ (all with $\epsilon=10^{-9}$),
(p)-(t) TaAs with convergence tolerance $\epsilon=10^{-7}$,
and (u)-(y) TaAs with $\epsilon=10^{-9}$.
}
\label{fig:dos}
\end{figure*}

Now, we investigate the effects of inaccurate stitching of the MLWFs on actual physical quantities calculated from the combined tight-binding model for the surface.
In Fig.~\ref{fig:dos}, we show the momentum-resolved bulk and surface density of states (DOS) calculated from the semi-infinite surface using the iterative method for obtaining
the Green function~\cite{Sancho1985}.

We find that the momentum-resolved surface DOS calculated from the bulk-derived surface model, i.\,e.\,, by using the surface tight-binding model constructed by using exactly the same on-site potentials and hopping integrals as the bulk, significantly deviates from
the momentum-resolved surface DOS calculated from slab-derived models.
This result demonstrates the importance of properly taking into account the deviation of the electronic structure at surfaces from that of the bulk in surface simulations.

Also, we find in Fig.~\ref{fig:dos} that for some materials, if no correction is applied, or if only the minimal correction is applied, bands that do not occur in case of the Hamiltonian correction appear in the momentum-resolved surface DOS. The black arrows in 
Figs.~\ref{fig:dos}(m), \ref{fig:dos}(w), and~\ref{fig:dos}(x) indicate these additional bands.
From these considerations, we conclude that the additional bands are non-physical impurity bands, created due to the erroneous stitching of the bulk and slab MLWFs.
These bands are a direct evidence
showing that improper stitching of
the bulk and slab MLWFs can lead to artifacts in the physical quantities calculated from the combined tight-binding model.

In the case of TaAs, we find that the non-physical bands occur only when $\epsilon \le 10^{-8}$.
Note that the occurrence of these bands are likely be related with the large jump in the average overlap error at $\epsilon=10^{-8}$ shown in Fig.~\ref{fig:conv_tol}(d). The non-physical bands are not removed after the minimal correction. On the contrary, the momentum-resolved surface DOS for $\epsilon=10^{-7}$ and $\epsilon=10^{-9}$ with the Hamiltonian correction are indistinguishable.
These results show that the Hamiltonian correction can effectively reduce the errors that occur during maximal localization, even when minimal correction fails to do so.

Using our methods, we have also generated the tight-binding models of diamond, GeTe, and Bi$_2$Se$_3$ slabs thicker than the ones used for the seamless stitching and found that their energy eigenvalues are in excellent agreement with those obtained from a direct DFT calculations on the corresponding thicker slabs (not shown).
This test demonstrates the validity of our methods.

\section{Comparison with the Projection-Only Wannier Functions}
Finally, we compare the Hamiltonian corrected MLWFs with the projection-only WFs. Projection-only WFs are constructed by projecting the initial guesses to the target subspace of Kohn-Sham eigenstates, and then applying
L\"owdin orthogonalization.
The target subspace can be determined either with~\cite{Zhang2010NJP} or without~\cite{ku2002prl,anisimov2005prb,Ku2010PRL,Berlijn2011PRL}
the disentanglement step, where the gauge-invariant part of $\Omega$ is iteratively minimized~\cite{Souza2001PRB}.
The gauge-variant part of $\Omega$ however is not further minimized.

Since the projection-only WFs retain the atomic-orbital-like
features of the initial guesses, projection-only WFs obtained from different systems tend to be similar.
Thus, they can be used as the basis functions to combine {\it ab initio} tight-binding models without further correction.

\begin{figure}
\includegraphics[width=0.95\columnwidth]{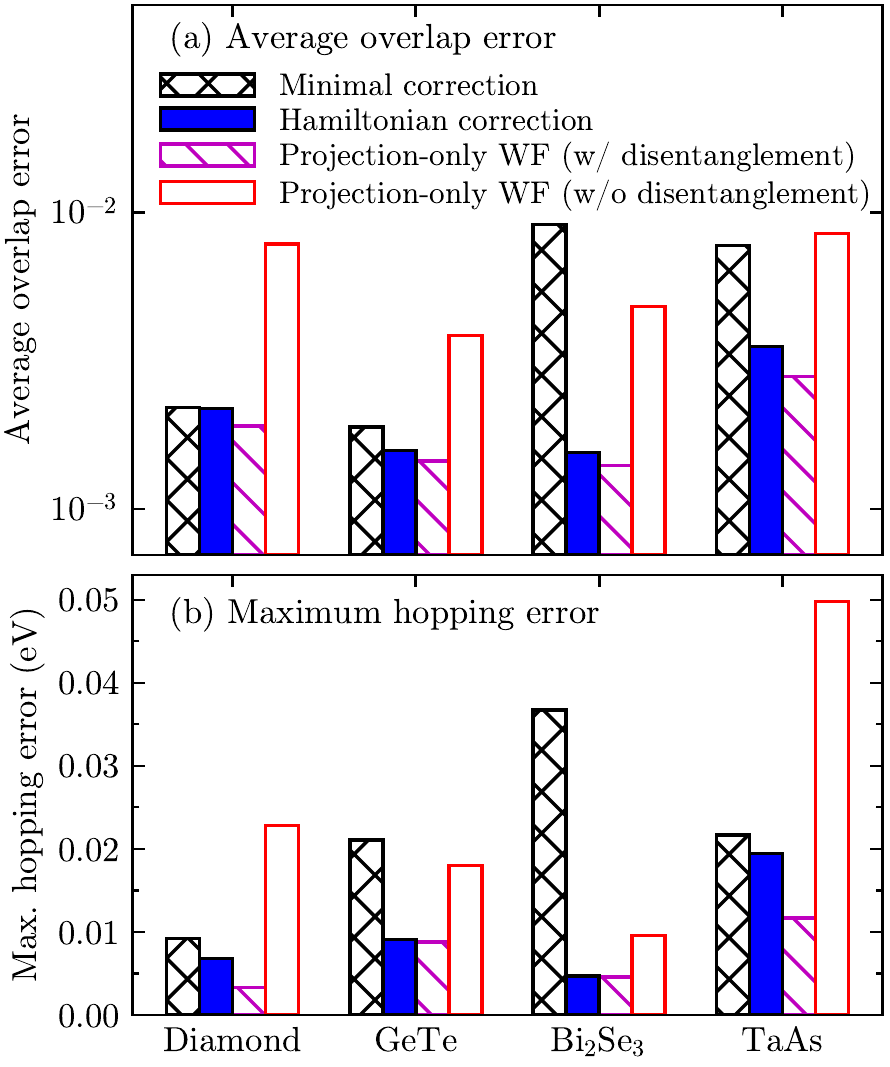}
\caption{
The (a) average overlap error and (b) maximum hopping error for different types of WFs.
For each material, the errors are calculated for MLWFs after
minimal or Hamiltonian correction and for projection-only WFs,
either with or without disentanglement.
}
\label{fig:proj_error}
\end{figure}

In Fig.~\ref{fig:proj_error}, we show the average overlap error and the maximal hopping error obtained from the corrected MLWFs and the projection-only WFs.
First of all, the projection-only WFs without disentanglement
result in much larger errors than those with disentanglement.
Therefore, for a fair comparison, we exclude the case of projection-only
WFs without disentanglement in the following discussion.
While the errors of the MLWFs after the minimal correction are much larger than the others, the errors after the Hamiltonian correction are comparable to those of the projection-only WFs. We note that the minimal and the Hamiltonian corrections
have no effect on the projection-only WFs.

\begin{figure}
\includegraphics[width=0.95\columnwidth]{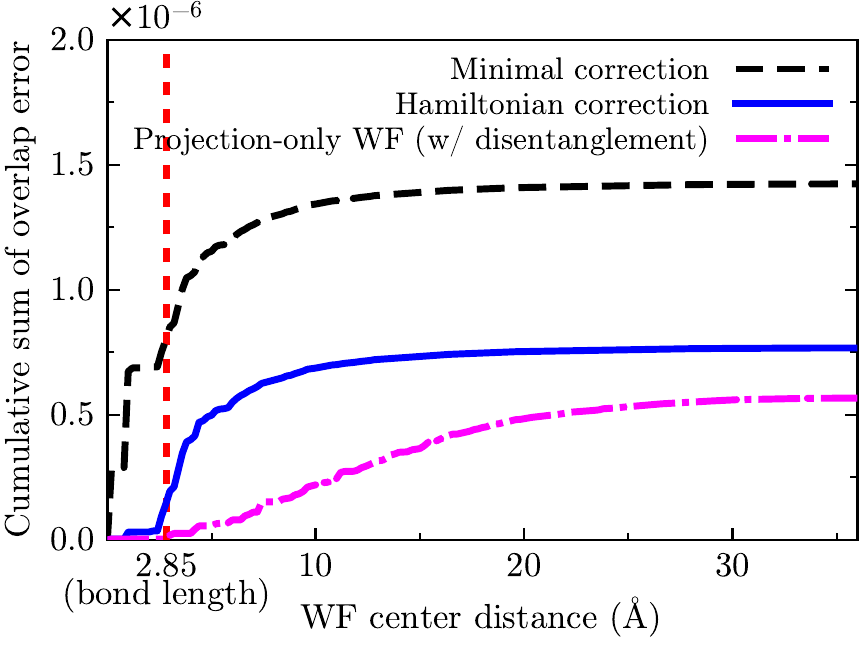}
\caption{
The cumulative sum of squared individual overlap errors with respect to the distance between the centers of the WFs for GeTe. The vertical dashed line indicates the length of the shortest Ge-Te bond. The cumulative sum is normalized so that its square root reaches the average overlap error shown in Fig.~\ref{fig:proj_error}(a).
}
\label{fig:overlap_cumul}
\end{figure}

For further investigation, we look for the dominant sources of the overlap errors. In Fig.~\ref{fig:overlap_cumul}, we show for GeTe the cumulative sum of individual overlap errors as a function of the distance between the centers of the WFs.
In the case of minimal correction, around half of the overlap error originates from
the WFs whose centers are closer than the bond length.
In other words, the nonorthogonality between the bulk and slab MLWFs centered on the same atom is the dominant source of error.

In the case of Hamiltonian correction, the nonorthogonality between MLWFs centered on the same atom is almost completely suppressed. Instead, overlap between MLWFs centered on nearby atoms are dominant.
The MLWFs whose centers are separated by longer
than 10~{\rm \AA} make negligible contribution to the average overlap error, reflecting the locality of MLWFs.

Contrary to the previous two cases, the overlap error between the projection-only WFs comes from a broad range of WF center distance,
from 3 to 20~{\rm \AA}.
The projection-only WFs are less localized than the MLWFs
and have non-negligible overlap even if the
centers of the WFs
are separated far apart.

An advantage of using projection-only WFs is that one can generate a tight-binding model
having the symmetry of the system, such as discrete rotational symmetry. However, one can obtain MLWF-based tight-binding models with the symmetry
of the electronic system by modifying the localization functional~\cite{Wang2014PRB} or by explicitly imposing the symmetry~\cite{Sakuma2013PRB}. Since the corrections proposed in this paper are post-processing methods,
one can apply them to these generalized
MLWFs to construct combined tight-binding models based on MLWFs that preserve the symmetry.

In summary, MLWFs with Hamiltonian correction are more localized than the projection-only WFs, while having comparable or slightly larger errors.
Therefore, by using MLWFs and the Hamiltonian correction we can seamlessly stitch two tight-binding models
without resorting to the similarity of the initial guesses for the bulk and the slab WFs. This combination of methods also enables to seamlessly stitch two tight-binding models obtained from the method of automatically generating MLWFs \cite{damle2017,damle2018}, in which the initial guesses for the WFs are not manually selected but are automatically constructed using only a few parameters.

In passing we note that obviously
one cannot use the Hamiltonian correction if the two systems to be combined have different atomic structures
in the common region where the WFs
in the two systems are combined.
This is because the Hamiltonian correction is based on the similarity of the Hamiltonian at the level of {\it ab initio} calculation. Use of WFs for band unfolding~\cite{Ku2010PRL}, virtual crystal approximation~\cite{Liu2013PRB}, and interpolation of SOC~\cite{Gotlieb2017PRB} or strain~\cite{Gresch2018PRM} belong to this category.

\section{Conclusion}
In conclusion, we proposed two post-processing methods that correct the difference between MLWFs obtained from different systems.
We tested our methods on the surfaces of four materials
and assessed the results based on the overlap error and the hopping error.
We showed that the minimal correction is successful in correcting the discrepancy between the MLWFs if the convergence criterion for the total spread is sufficiently tight,
and the orbital wavefunctions of the bulk and slab MLWFs do not significantly
differ from each other.
However, depending on the material and on
the convergence criterion, the minimal correction resulted in
large overlap and hopping errors in some cases.
On the other hand, the
Hamiltonian correction was
found to be much more accurate and generally applicable.
The errors of the Hamiltonian-corrected MLWFs were comparable to that of the projection-only WFs.
These corrections can be easily used in the study of surfaces, interfaces, and defects to obtain reliable results from the combined {\it ab initio} tight-binding models based on MLWFs.

\begin{acknowledgments}
This work was supported by the Creative-Pioneering Research Program through Seoul National University.
\end{acknowledgments}

\appendix
\section{Plane-wave basis methods}
In this appendix, we detail the methods specific to plane-wave basis DFT calculations.
Using a plane-wave basis set defined as
\begin{equation} \label{eq:appa_pw}
    \braket{\mathbf{r}}{\mathbf{k+G}} = \frac{1}{\sqrt{V_{\rm cell}}} e^{i (\mathbf{k+G}) \cdot \mathbf{r}}  
\end{equation}
where $V_{\rm cell}$ is the volume of the unit cell, we can represent the Kohn-Sham eigenstates of the bulk and the slab as
\begin{equation}
\ket{\psi_{j\mathbf{k'}}^{\rm B}} = \sum_{\mathbf G'}
c_{j\mathbf{k'G'}}^{\rm B} \ket{\mathbf{k'+G'}}
\label{eq:appa_psiB}
\end{equation}
and
\begin{equation}
\ket{\psi_{n\mathbf{k}}^{\rm S}} = \sum_{\mathbf G} c_{n\mathbf{kG}}^{\rm S} \ket{\mathbf{k+G}}\,,
\label{eq:appa_psiS}
\end{equation}
respectively.
Here, $\mathbf{k'}$ and $\mathbf{G'}$ are the Bloch wavevector and the reciprocal-lattice vector of the bulk, respectively, and $\mathbf{k}$ and $\mathbf{G}$ are the corresponding quantities of the slab. Also, $j$ and $n$ are the band indices of the bulk and slab Bloch states, respectively.

To calculate the inner product between the bulk and slab wavefunctions, we choose the
plane-wave bases of the bulk and the slab to be commensurate. Concretely, let the
fast fourier transformation (FFT) grid 
for wavefunctions be $N_1 \times N_2 \times N_3$
and the k-point grid for Wannierization be $n_1 \times n_2 \times n_3$ for the bulk.
Then, we choose the slab supercell to have the same in-plane lattice parameters
as the bulk and set the out-of-plane
lattice parameter as $n_3$ times
the out-of-plane lattice parameter of the bulk. The FFT grid for DFT calculation and the k-point grid for Wannierization of the slab
are chosen as $N_1 \times N_2 \times N_3 n_3$
and $n_1 \times n_2 \times 1$, respectively. With these choices, the inner product between the bulk and slab wavefunctions can be calculated using the inner product of the bulk and slab plane waves,
\begin{equation} \label{eq:appa_pwortho}
    \braket{\mathbf{k+G}}{\mathbf{k'+G'}} = \delta_{\mathbf{k+G},\mathbf{k'+G'}}\,.
\end{equation}

The unitary transform from the Kohn-Sham eigenstates to the MLWFs is given as
\begin{equation}
\ket{w_{\mathbf{R'}i}^{\rm B}} 
= \frac{1}{\sqrt{N_{\rm W}^{\rm B}}}\sum_{{\mathbf k'},\,j} e^{-i\mathbf{k'} \cdot \mathbf{R'}} \ket{\psi_{j\mathbf{k'}}^{\rm B}} \left(U_{{\rm B}}^{\mathbf{k'}}\right)_{ji}
\label{eq:appa_wB}
\end{equation}
and
\begin{equation}
\ket{w_{\mathbf{R}m}^{\rm S}} 
= \frac{1}{\sqrt{N_{\rm W}^{\rm S}}}\sum_{{\mathbf k},\,n}
e^{-i\mathbf{k} \cdot \mathbf{R}}\ket{\psi_{n\mathbf{k}}^{\rm S}}
\left(U_{{\rm S}}^{\mathbf{k}}\right)_{nm}\,.
\label{eq:appa_wS}
\end{equation}

Now, we calculate the overlap matrix $A$ defined in Eqs.~\eqref{eq:overlap_amat1}. From Eqs.~\eqref{eq:appa_wB} and~\eqref{eq:appa_wS},
one obtains
\begin{eqnarray}
    \label{eq:ARm0u}
    &&A_{\mathbf{R}m,\mathbf{0}i}
    = \braket{w_{\mathbf{R}m}^{\rm S}}{w_{\mathbf{0}i}^{\rm B}} \label{eq:appa_ar} \\
    =&& \frac{1}{\sqrt{N_{\rm W}^{\rm B}N_{\rm W}^{\rm S}}}
    \sum_{\mathbf{k,\,k'},\,n,\,j} e^{i\mathbf{k} \cdot \mathbf{R}}
    \left(U_{{\rm S}}^{\mathbf{k}\dagger}\right)_{mn} \braket{\psi_{n\mathbf{k}}^{\rm S}}{\psi_{j\mathbf{k'}}^{\rm B}}
    \left(U_{{\rm B}}^{\mathbf{k'}}\right)_{ji}. \nonumber
\end{eqnarray}
From Eqs.~\eqref{eq:overlap_amat2}, \eqref{eq:appa_pwortho},
and~\eqref{eq:ARm0u}, one obtains
\begin{eqnarray}
    \label{eq:Akmi}
    &&A^{\mathbf{k}}_{mi}
    = \sum_{\mathbf{R}} e^{-i\mathbf{k}\cdot\mathbf{R}} A_{\mathbf{R}m,\mathbf{0}i} \nonumber \\
    =&& \sqrt{\frac{N_{\rm W}^{\rm S}}{N_{\rm W}^{\rm B}}}
    \sum_{\mathbf{k'},\,n,\,j} \left(U_{{\rm S}}^{\mathbf{k}\dagger}\right)_{mn} \braket{\psi_{n\mathbf{k}}^{\rm S}}{\psi_{j\mathbf{k'}}^{\rm B}}
    \left(U_{{\rm B}}^{\mathbf{k'}}\right)_{ji} \label{eq:appa_ak} \\
    =&& \sqrt{\frac{N_{\rm W}^{\rm S}}{N_{\rm W}^{\rm B}}}
    \sum_{\substack{\mathbf{k',G,G'}\\n,\,j}}
    \left(U_{{\rm S}}^{\mathbf{k}\dagger}\right)_{mn}
    c_{n\mathbf{kG}}^{\rm{S}*} c_{j\mathbf{k'G'}}^{\rm B} 
    \left(U_{{\rm B}}^{\mathbf{k'}}\right)_{ji}
    \delta_{\mathbf{k+G},\mathbf{k'+G'}}\,. \nonumber
\end{eqnarray}
Note that the sum over ${\mathbf k'}$ is calculated only over those having the same in-plane component as ${\mathbf k}$, since otherwise $\braket{\psi_{n\mathbf{k}}^{\rm S}}{\psi_{j\mathbf{k'}}^{\rm B}}=0$.
From Eq.~\eqref{eq:Akmi}, the overlap matrix $A_{mi}^{\mathbf k}$ can be calculated from the $c_{\mathbf{G}}$ coefficients and the $U$ matrices, which are obtained from the {\it ab initio} calculation and Wannierization, respectively.

Next, we explain the calculation of WF signatures
used in the minimal correction.
WF signatures are the coefficients of the WFs in the plane-wave basis, as defined in Eq.~\eqref{eq:mincorr_sig}.
It can be straightforwardly calculated using Eqs.~\eqref{eq:appa_pw}, \eqref{eq:appa_psiS}, and~\eqref{eq:appa_wS} for the slab,
\begin{eqnarray}
    I_{m}^{\rm S}(\mathbf{G})
    &=& \frac{1}{\sqrt{V_{\rm cell}}}\int_{V_{\rm cell}} d\mathbf{r} e^{-i \mathbf{G}\cdot(\mathbf{r}-\mathbf{r_{\rm c}})} \braket{\mathbf r}{w_{\mathbf{0}m}} \nonumber \\
    &=& e^{i \mathbf{G}\cdot \mathbf{r_{\rm c}}} \braket{\mathbf{G}}{w_{\mathbf{0}m}} \nonumber \\
     &=& \frac{1}{\sqrt{N_{\rm W}^{\rm S}}} e^{i \mathbf{G}\cdot \mathbf{r_{\rm c}}} \sum_{\mathbf k}  \braket{\mathbf{G}}{\psi_{n\mathbf{k}}^{\rm s}}\, \left(U_{{\rm S}}^{\mathbf{k}}\right)_{nm} \nonumber \\
     &=& \frac{1}{\sqrt{N_{\rm W}^{\rm S}}} e^{i \mathbf{G}\cdot \mathbf{r_{\rm c}}}\, \sum_n\,c_{n\mathbf{0G}}^{\rm S}\, \left(U_{{\rm S}}^{\mathbf{0}}\right)_{nm}\,,
     \label{eq:appa_sigs}
\end{eqnarray}
and analogously for the bulk,
\begin{equation} \label{eq:appa_sigb}
    I_{i}^{\rm B}(\mathbf{G'})
     = \frac{1}{\sqrt{N_{\rm W}^{\rm B}}} e^{i \mathbf{G'}\cdot \mathbf{r_{\rm c}}}\, \sum_j\,c_{j\mathbf{0G'}}^{\rm B} (U_{{\rm B}}^{\mathbf{0}})_{ji}\,.
\end{equation}
Note that only the plane-wave coefficients for $\mathbf{k}=\mathbf{0}$ is used.

If the out-of-plane lattice constant of the slab supercell is not an integer multiple of that of the bulk, one cannot compare the bulk and slab WF signatures corresponding to a
G vector with nonzero out-of-plane component.
Since our minimal correction aims at working
irrespective
of the supercell structure,
we do not use signatures at G vectors
with nonzero out-of-plane component during corrections.

In practice, we use the WF signatures at five G vectors: ${\bf G}$=(0,0,0), ($\pm$1,0,0), and (0,$\pm$1,0), in units of
reciprocal lattice vectors.
We find these G vectors
sufficient for all the materials we have tested,
but it is possible to increase the number of
G vectors if necessary.
Since the WF signatures are calculated for only a few G vectors,
the required computational cost is negligible.

\section{Steepest descent minimization algorithm}
In this appendix, we explain the steepest descent algorithm we use for the Hamiltonian correction.
The Hamiltonian correction is achieved by finding the optimal $V^{(0)}$ matrix,
common to all ${\mathbf k}$, that minimizes the square of the average hopping error
\begin{equation} \label{eq:appb_obj}
    \Delta H_{\rm ave}^2 \left[V^{(0)}\right]
    = \frac{1}{N_{\rm W}^{\rm B}N_{\rm k}^{\rm S}} \sum_{\mathbf k} \fnorm{ V^{(0) \dagger} H_{\rm S}^{\mathbf k} V^{(0)} - H_{\rm B}^{\mathbf k}}^2
\end{equation}
under the constraint that $V^{(0)}$ is a block-diagonal
unitary matrix. Matrices $H_{\rm S}^{\mathbf k}$ and $H_{\rm B}^{\mathbf k}$ are square
hermitian matrices of dimension $N_{\rm W}^{\rm B}$.

Hereafter in this appendix, we omit the superscript $(0)$ in $V^{(0)}$ for brevity. Omitting the irrelevant factor $1 / N_{\rm W}^{\rm B}N_{\rm k}^{\rm S}$ from Eq.~\eqref{eq:appb_obj} and using the unitarity of $V$, one can show that
\begin{eqnarray}  \label{eq:appb_obj2}
    &&J[V] = N_{\rm W}^{\rm B}N_{\rm k}^{\rm S} \Delta H_{\rm ave}^2[V] \nonumber \\
    =&& \sum_{\mathbf k} \left( \fnorm{H_{\rm S}^{\mathbf k}}^2 + \fnorm{H_{\rm B}^{\mathbf k}}^2 - 2 \Tr V^\dagger H_{\rm S}^{\mathbf k} V H_{\rm B}^{\mathbf k} \right)
\end{eqnarray}
is the cost function that should be minimized.
Since the first two terms in the parentheses do not depend on $V$, only the last term need to be considered during minimization.

To exploit the block-diagonality constraint on $V$, we
define the projection operator
$P_a$, which projects onto the subspace of MLWFs centered at
the atom labelled with $a$.
One can write $V$ as a sum of blocks
\begin{equation} \label{eq:appb_va}
    V = \sum_a P_a V P_a = \sum_a V_a\,,
\end{equation}
where we define $V_a = P_a V P_a$ as the diagonal block of $V$. Due to the unitarity of $V$, $V_a$ is also a unitary matrix in the $P_a$ subspace. We simultaneously optimize all $V_a$ matrices using the steepest descent method.

The algorithm we implement is the ``self-tuning Riemannian steepest descent
algorithm,''
as summarized in Table~II of Ref.~\cite{abrudan2008steepest}. This algorithm is
suitable
for our purpose since the unitarity of $V$ is explicitly maintained during the minimizaion steps. Hence, additional orthogonalization of the output is not needed.

In the remaining part of the appendix, we state formulae used in the algorithm. First, the gradient of the cost function is defined as
\begin{equation} \label{eq:appb_grad}
    (\Gamma_a)_{ij}
    = \frac{\partial J}{\partial (V_a)_{ij}^*}
    = -2 \sum_{\mathbf k} (P_a H_{\rm S}^{\mathbf k} V H_{\rm B}^{\mathbf k} P_a)_{ij}\,.
\end{equation}
The gradient direction on the Riemannian space is defined as
\begin{equation} \label{eq:appb_rgrad}
    G_a = \Gamma_a V_a^\dagger - V_a \Gamma_a^\dagger\,.
\end{equation}
By definition, $G_a$ is an anti-hermitian matrix.

The $V_a$ matrix is updated by multiplying it by the update matrix
$Q_a$ defined as
\begin{equation} \label{eq:appb_pa}
    Q_a = \exp(-\mu G_a)\,,
\end{equation}
with a step size parameter $\mu$. The step size $\mu$ is a positive real number that is adaptively updated by multiplying or dividing by 2. Matrix $Q_a$ is unitary since it is an exponential of an anti-hermitian matrix $G_a$. Hence, the updated matrix $Q_a V_a$ remains unitary. The set of matrices $Q_a V_a$ is used as
the input $V_a$ for the next iteration.

To determine convergence of iterations, we use the sum of the squared norm of the Riemannian gradient
\begin{equation} \label{eq:appb_norm}
    {\cal N} = \frac{1}{2} \sum_a \Tr (G_a G_a^\dagger)\,.
\end{equation}
The iteration is assumed to converge when ${\cal N} < N_{\rm W}^{\rm B} N_{\rm k}^{\rm S} \times 10^{-5} \rm{~eV^2}$ is satisfied.
In all cases we tested, at most a few tens of iterations were sufficient to reach convergence.

\begin{figure}
\includegraphics[width=0.95\columnwidth]{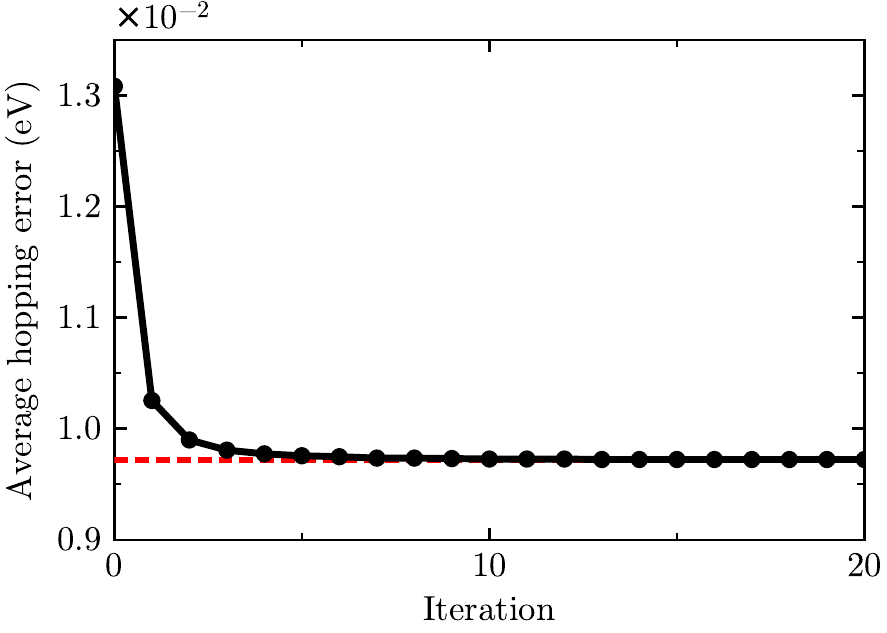}
\caption{
The average hopping error verses the
number of steepest descent iterations
for the Hamiltonian correction applied to GeTe.
The horizontal dashed line indicates the converged value.
}
\label{fig:minimization}
\end{figure}

We show in Fig.~\ref{fig:minimization} that the average hopping error, which is proportional to the square root of the objective functional $J$, monotonically decreases at each iteration.

\bibliography{main}

\end{document}